\documentclass[11pt,a4paper]{article}%
\usepackage{amsfonts}
\usepackage{amsmath}
\usepackage{amssymb}
\usepackage[margin=1in,a4paper]{geometry}
\usepackage[sc,center,compact,noindentafter,medium]{titlesec}
\usepackage[onehalfspacing,nodisplayskipstretch]{setspace}
\usepackage{graphicx}
\usepackage{color}%
\providecommand{\U}[1]{\protect\rule{.1in}{.1in}}
\graphicspath{{C:/temp/}}

\definecolor{darkblue}{rgb}{.2, 0.2,.8}
\definecolor{darkgreen}{rgb}{0,0.5,0.3}
\definecolor{darkred}{rgb}{.8, .1,.1}

\providecommand{\U}[1]{\protect\rule{.1in}{.1in}}

\newtheorem{theorem}{\normalfont\scshape Theorem}[section]
\newtheorem{corollary}{\normalfont\scshape Corollary}[section]
\newtheorem{lemma}{\normalfont\scshape Lemma}[section]

\newtheorem{remark}{\normalfont\scshape Remark}[section]

\expandafter
\def \expandafter \normalsize \expandafter{\normalsize \setlength \abovedisplayskip{10pt plus 2pt minus 7pt}}
\expandafter
\def \expandafter \normalsize \expandafter{\normalsize \setlength \abovedisplayshortskip{0pt plus 2pt}}
\expandafter
\def \expandafter \normalsize \expandafter{\normalsize \setlength \belowdisplayskip{10pt plus 2pt minus 7pt}}
\expandafter
\def \expandafter \normalsize \expandafter{\normalsize \setlength \belowdisplayshortskip{5pt plus 2pt minus 3pt}}

\numberwithin{equation}{section}
\begin{document}

\title{ }

\begin{center}
{\LARGE \textsc{Beyond the Mean: Limit Theory and Tests for Infinite-Mean Autoregressive Conditional Durations}}%

\renewcommand{\thefootnote}{}
\footnote{
\hspace{-7.2mm}
$^{a}%
$Department of Economics, University of Bologna, Italy and Department of Economics, University of Exeter, UK.
\newline$^{b}%
$Department of Mathematical Sciences, University of Copenhagen, Denmark.
\newline$^{c}$Department of Economics, University of Copenhagen, Denmark.
\newline
A.~Rahbek and G.~Cavaliere gratefully acknowledge support from the Independent Research Fund Denmark (DFF Grant 7015-00028) and
the Italian Ministry of University and Research (PRIN 2020 Grant 2020B2AKFW).
The paper was presented at the Zaragoza time series workshop, April 2025, and we thank participants there for comments.
We also thank Stefan Voigt for providing the code used to obtain the duration data analyzed in our empirical application;
see also https://www.tidy-finance.org/ for more information. Correspondence to: Anders Rahbek, Department of Economics, University of Copenhagen, email anders.rahbek@econ.ku.dk.
}
\addtocounter{footnote}{-1}
\renewcommand{\thefootnote}{\arabic{footnote}}%
{\normalsize \vspace{0.1cm} }

{\large \textsc{Giuseppe Cavaliere}}$^{a}${\large \textsc{, Thomas Mikosch}%
}$^{b}$, {\large \textsc{Anders Rahbek}}$^{c}$

{\large \textsc{and Frederik Vilandt}}$^{c}${\normalsize \vspace{0.2cm}%
\vspace{0.2cm}}

May 9, 2025{\normalsize \vspace{0.2cm}\vspace{0.2cm}}

\bigskip

\textsc{Abstract}\vspace{-0.15cm}
\end{center}

{\small Integrated autoregressive conditional duration (ACD) models serve as
natural counterparts to the well-known integrated GARCH models used for
financial returns. However, despite their resemblance, asymptotic theory for
ACD is challenging and\ also not complete, in particular for integrated ACD.
Central challenges arise from the facts that (i) integrated ACD processes
imply durations with infinite expectation, and (ii) even in the non-integrated
case, conventional asymptotic approaches break down due to the randomness in
the number of durations within a fixed observation period. Addressing these
challenges, we provide here unified asymptotic theory for the (quasi-) maximum
likelihood estimator for ACD\ models; a unified theory which includes
integrated ACD models. Based on the new results, we also provide a novel
framework for hypothesis testing in duration models, enabling inference on a
key empirical question: whether durations possess a finite or infinite
expectation.}

{\small We apply our results to high-frequency cryptocurrency ETF\ trading
data. Motivated by parameter estimates near the integrated ACD boundary, we
assess whether durations between trades in these markets have finite
expectation, an assumption often made implicitly in the literature on point
process models. Our empirical findings indicate infinite-mean durations for
all the five cryptocurrencies examined, with the integrated ACD hypothesis
rejected -- against alternatives with tail index less than one -- for four out
of the five cryptocurrencies considered.}

\bigskip

\bigskip


\bigskip

\newpage

\section{Introduction}

The recent work by Cavaliere, Mikosch, Rahbek, and Vilandt~(2024, 2025)
introduce a novel non-standard asymptotic theory for (quasi-) maximum
likelihood estimators ((Q)MLE) for stationary and ergodic autoregressive
conditional duration (ACD) models. Prior to these contributions, estimation
and inference in ACD models were assumed to follow from standard asymptotic
theory; see, e.g., Engle and Russell~(1998), Bhogal and Variyam Thekke~(2019),
Fernandes, Medeiros and Veiga~(2016), Hautsch~(2011) and Saulo, Pal, Souza,
Vila and Dasilva~(2025).

A key challenge for the estimation theory in ACD\ models is that the number of
durations, and hence observations $n(t)$, for a given time span $[0,t]$,
$t>0$, is random. The randomness of the number of observations $n(t)$ implies
that classical limit results, including standard laws of large numbers and
central limit theorems, cannot be applied directly, or even applied at all, as
they rely on the assumption of a deterministically increasing number of
observations. A main implication of the randomness of $n(t)$ is that a crucial
role is played by the tail index $\kappa\in(0,\infty)$ of the marginal
distribution of the stationary and ergodic durations $x_{i}$, defined by the
condition $\mathbb{P}(x_{i}>z)\sim c_{\kappa}z^{-\kappa}$ as $z\rightarrow
\infty$ for some $c_{\kappa}>0$. To deal with this, new non-standard theory
was developed in Cavaliere et al.~(2024, 2025).

In short, the tail index $\kappa$ of the durations determines both \emph{(i)
}the rate of convergence of the likelihood-based estimators, as well as
\emph{(ii)} the limiting distribution of these. More precisely, Cavaliere,
Mikosch, Rahbek, and Vilandt~(2025), henceforth CMRV, demonstrate that for
$0<\kappa<1$, durations have infinite mean and the limiting distribution of
the estimators is mixed normal with convergence rate $\sqrt{t^{\kappa}}$. On
the other hand, when $\kappa>1$, the durations have finite expectation, and
asymptotic normality holds for the estimators at the standard $\sqrt{t}$-rate.
Notably, the empirically relevant case of $\kappa=1$ is excluded in CMRV;\ as
stated in their Remark 5, for $\kappa=1$ `\emph{the limiting behavior [of the
estimators] is unknown}'.

In this paper we extend the theory to include the case of $\kappa=1$ as well,
and enabling us to provide a unifying framework for estimation and asymptotic
inference in ACD\ models.

In terms of the classical ACD process for durations, or waiting times,
$x_{i}>0$, formally introduced in\ Section \ref{sec ACD model} below, the
points above can be summarized as follows. Consider $x_{i}=\psi_{i}%
\varepsilon_{i}$, $\varepsilon_{i}$ i.i.d. with $\mathbb{E}\left[
\varepsilon_{i}\right]  =1$, and conditional duration $\psi_{i}$ given by%
\begin{equation}
\psi_{i}=\omega+\alpha x_{i-1}+\beta\psi_{i-1}\text{,\qquad}i=1,2,\ldots
,n(t)\text{.} \label{eq cond int 1}%
\end{equation}
In terms of the parameters $\alpha,\beta>0$ the results in CMRV include
$\alpha+\beta>1$ (i.e. $\kappa<1$) and $\alpha+\beta<1$ (i.e. $\kappa>1$),
while the empirically relevant case of $\alpha+\beta=1$, and hence $\kappa=1$,
is excluded.

Note that for the well-known classical generalized autoregressive
heteroskedastic model (GARCH), the conditional variance has a functional form
in terms of the parameters $\omega,\alpha$ and $\beta$, identical to that of
the conditional duration, or (inverse) intensity, $\psi_{i}$ of the ACD
process in~(\ref{eq cond int 1}). Hence, by analogy with the typically
witnessed `integrated GARCH'\ case of $\alpha+\beta=1$ when modelling
financial returns, we label this case \emph{integrated ACD }(IACD). While the
conditional duration in an ACD model resembles that of the conditional
variance in the GARCH model, and hence their likelihood functions share key
structures, the asymptotic theory for the likelihood estimators are very
different. Thus, in contrast to the ACD, GARCH likelihood estimators are
asymptotically Gaussian at a standard rate of convergence, regardless of
whether $\alpha+\beta\neq1$ or $\alpha+\beta=1$.

We complete here the estimation theory for ACD\ models by showing that for
$\kappa=1$, the rate of convergence for the quasi maximum likelihood estimator
(QMLE) is the non-standard $\sqrt{t/\log t}$-rate, and the limiting
distribution Gaussian. The novel result and theory, in combination with the
results in CMRV allow us to develop formal testing of the hypothesis of IACD
in combination with testing infinite mean against finite mean. The latter may
be stated as testing $\alpha+\beta\geq1$, against the alternative
$\alpha+\beta<1$ ($\kappa\leq1$ against $\kappa>1$). Notice that this idea is
reminiscent of testing for finite variance in strictly stationary
GARCH\ processes as in Francq and Zako\"{\i}an~(2022), in
double-autoregressive models as in Ling~(2004) and in non-causal stationary
autoregressions as in Gourierox and Zako\"{\i}an~(2017). As mentioned, this is
of key interest in applications as most often the sum of the estimators of
$\alpha$ and $\beta$ is close to one, similar to estimation in GARCH models.

The results are illustrated by likelihood analyses on recent high-frequency
trade data for exchange traded funds (ETFs) on cryptocurrencies. We find that
the ACD model provides estimators of $\alpha\,$and $\beta$ summing
approximately to one. However, while $\alpha+\beta\simeq1$, based on
implementation of the testing results, we do not reject $\alpha+\beta\geq1$,
implying a tail index $\kappa\leq1$, in line with what might be expected for
cryptocurrencies due to their highly irregular trading patterns.

The paper is structured as follows. In Section~\ref{sec ACD model} we
introduce the ACD model and derive its key feature in the IACD\ case. In
Section~\ref{Sec main} we present the asymptotic theory for the estimators and
the associated test statistics. In Section \ref{sec MC} we analyze the finite
sample properties of estimators and tests by Monte Carlo simulation. The
empirical analysis of cryptocurrencies is presented in Section
\ref{sec empirical}. Section \ref{sec conc} concludes. All proofs are provided
in the Appendix.

\section{The ACD Model}

\label{sec ACD model}

Engle and Russell~(1998) proposed the ACD model in order to analyze $n(t)$
observations of waiting times, or durations, $\left\{  x_{i}\right\}
_{i=1}^{n(t)}$, within a time span $[0,t]$, $t>0$, which can be a day, a year
or some other pre-specified observation period. The ACD has a multiplicative
form and is given by%
\begin{equation}
x_{i}=\psi_{i}(\theta)\varepsilon_{i}\text{, \ \ \ }i=1,...,n(t),
\label{eq ACD model x}%
\end{equation}
where $\varepsilon_{i}$ is an i.i.d. sequence of strictly positive random
variables with $\mathbb{E}\left[  \varepsilon_{i}\right]  =1$ and
$\mathbb{V[}\varepsilon_{i}]=\sigma_{\varepsilon}^{2}<\infty$, and with
density bounded away from zero on compact subsets of $\mathbb{R}_{+}$.
Moreover, the conditional duration is given by (\ref{eq cond int 1}), or%
\begin{equation}
\psi_{i}(\theta)=\omega+\alpha x_{i-1}+\beta\psi_{i-1}(\theta),
\label{eq ACD model int}%
\end{equation}
in terms of the parameter vector $\theta=\left(  \omega,\alpha,\beta\right)
^{\prime}\in\mathbb{R}_{+}^{3}$. \ With $\Theta$ a compact subset of
$\mathbb{R}_{+}^{3}$, the QMLE $\hat{\theta}_{t}$ is defined by
\begin{equation}
\hat{\theta}_{t}=\arg\max_{\theta\in\Theta}\mathcal{L}_{n(t)}(\theta)\text{,}
\label{eq def QMLE U}%
\end{equation}
where $\mathcal{L}_{n(t)}(\theta)$ is the exponential log-likelihood
function,
\begin{equation}
\mathcal{L}_{n(t)}(\theta)=\sum_{i=1}^{n(t)}\mathcal{\ell}_{i}(\theta)\text{,
\ \ \ }\mathcal{\ell}_{i}(\theta)=-\left(  \log\psi_{i}(\theta)+x_{i}/\psi
_{i}(\theta)\right)  \text{,} \label{eq def Likelihood}%
\end{equation}
with initial values $x_{0}$ and $\psi_{0}\left(  \theta\right)  $.

Note that, as reflected by the definition of the log-likelihood in
(\ref{eq def Likelihood}), properties of both the random waiting times $x_{i}$
and, importantly, the corresponding random number of observations $n(t)$, are
key to the analysis of the QMLE, and these are therefore considered next.

\subsection{Properties of waiting times and number of observations}

With $\theta_{0}$ denoting the true parameter, it follows by CMRV that
$\left\{  x_{i}\right\}  $ in (\ref{eq ACD model x})\ is strictly stationary
and ergodic provided $\mathbb{E}\left[  \log\left(  \alpha_{0}\varepsilon
_{i}+\beta_{0}\right)  \right]  <0$ holds. Moreover, it holds that $x_{i}$ has
tail index $\kappa_{0}\in\left(  0,\infty\right)  $ given by the unique and
positive solution to
\begin{equation}
\mathbb{E}\left[  \left(  \alpha_{0}\varepsilon_{i}+\beta_{0}\right)
^{\kappa}\right]  =1. \label{eq kappa}%
\end{equation}
Recall here that if $x_{i}$ has tail index $\kappa_{0}$, then $\mathbb{E}%
\left[  x_{i}^{s}\right]  <\infty$, for positive $s<\kappa_{0}$, while
$\mathbb{E}\left[  x_{i}^{s}\right]  =\infty$ for $s\geq\kappa_{0}$.

By CMRV, if $0<\alpha_{0}+\beta_{0}<1$, $x_{i}$ is stationary and ergodic with
tail index $\kappa_{0}>1,$ such that $x_{i}$ has finite mean, $\mathbb{E}%
\left[  x_{i}\right]  =\mu_{0}=\omega_{0}\left(  1-\left(  \alpha_{0}%
+\beta_{0}\right)  \right)  ^{-1}<\infty$. On the other hand, if $\alpha
_{0}+\beta_{0}>1$ and $\mathbb{E}\left[  \log\left(  \alpha_{0}\varepsilon
_{i}+\beta_{0}\right)  \right]  <0$, $x_{i}$ is stationary and ergodic but
with tail index $\kappa_{0}<1$, and hence infinite mean, $\mathbb{E}\left[
x_{i}\right]  =\infty$.

Our focus here is on the yet unexplored case of integrated ACD where
$\alpha_{0}+\beta_{0}=1$ and $\mathbb{E}\left[  \log\left(  \alpha
_{0}\varepsilon_{i}+\beta_{0}\right)  \right]  <0$, and thus the case where
$x_{i}$ has tail index $\kappa_{0}=1$ and infinite mean.

It is central to the derivations of the asymptotic behavior of the
QMLE$\ \hat{\theta}_{t}$ that not only does the number of observations $n(t)$
increase as the observation span $[0,t]$ increases, or, equivalently as
$t\rightarrow\infty$, but also at which rate. That is, the results depend on
whether the number of observations in $[0,t]$, $n(t)$, appropriately
normalized by some positive increasing deterministic function of $t$, is
constant, or random, in the limit. Note in this respect that for a given time
span $\left[  0,t\right]  $, by definition $n(t)=\arg\max_{k\geq1}\{\sum
_{i=1}^{k}x_{i}\leq t\}$, and hence, by construction,
\begin{equation}
\sum_{i=1}^{n(t)}x_{i}\simeq t. \label{eq sum x_i approx t}%
\end{equation}
CMRV establish that for $\kappa_{0}>1$, the number of observations $n(t)$ and
$t$ are proportional in the sense that, $n(t)/t\overset{p}{\rightarrow}%
1/\mu_{0}$ as $t\rightarrow\infty$. In contrast,
$n(t)/t\overset{p}{\rightarrow}0$ for $\kappa_{0}<1$, and instead%
\begin{equation}
n(t)/t^{\kappa_{0}}\rightarrow_{d}\lambda_{\kappa_{0}},
\label{eq n over t kappa}%
\end{equation}
where $\lambda_{\kappa_{0}}$ is a strictly positive random variable. These
rates are reflected in Theorems 2 and 3 in CMRV which provide the limiting
distributions of $\hat{\theta}_{t}$ for $\kappa_{0}>1$ and $\kappa_{0}<1$,
respectively. Specifically, for $\kappa_{0}>1$, the QMLE satisfies that
$\sqrt{t}(\hat{\theta}_{t}-\theta_{0})$ is asymptotically Gaussian, while for
$\kappa_{0}<1$, with $\hat{\theta}_{t}$ the MLE, $\sqrt{t^{\kappa_{0}}}%
(\hat{\theta}_{t}-\theta_{0})$ is asymptotically mixed Gaussian.

For the case $\kappa_{0}=1$, as for the $\kappa_{0}<1$ case, it follows by
Lemma \ref{lem n div by g} below that $n(t)/t\overset{p}{\rightarrow}0$. That
is, when $\kappa_{0}\leq1$, the number of events $n(t)$ increase, but at a
slower pace than $t$. To see this intuitively, use (\ref{eq sum x_i approx t})
such that, as $t\rightarrow\infty$,
\[
n(t)/t\simeq%
\Big(%
\sum_{i=1}^{n(t)}x_{i}/n(t)%
\Big)%
^{-1}\rightarrow_{p}0\text{,}%
\]
which follows by using that for deterministic sequence $n$, $\frac{1}{n}%
\sum_{i=1}^{n}x_{i}$ diverges, since $\mathbb{E}\left[  x_{i}\right]  =\infty$
when $\kappa_{0}\leq1$.

As demonstrated in Lemma \ref{lem n div by g}, $n(t)$ normalized by $t/\log t$
has a constant limit in the integrated case. The proof of Lemma
\ref{lem n div by g} is given in the appendix and is based on results in
Buraczewski, Damek and Mikosch~(2016) and Jakubowski and Szewczak~(2021),
together with arguments from CMRV.

\begin{lemma}
\label{lem n div by g}Consider the ACD process for $x_{i}>0$ as given by
(\ref{eq ACD model x})-(\ref{eq ACD model int}) with $\theta_{0}=\left(
\omega_{0},\alpha_{0},\beta_{0}\right)  ^{\prime}$ satisfying $\omega
_{0},\alpha_{0}>0$ and the IACD\ hypothesis $\beta_{0}=1-\alpha_{0}\geq
0$\textbf{.} It follows that $x_{i}$ is stationary and ergodic with tail index
$\kappa_{0}=1$. Moreover,
\[
\frac{n(t)\log t}{t}\overset{p}{\rightarrow}1/c_{0}\text{, as }t\rightarrow
\infty\text{,}%
\]
with the constant $c_{0}\in(0,\infty)$ given by%
\begin{equation}
c_{0}=\omega_{0}\left(  {\mathbb{E}}\left[  \left(  1+\alpha_{0}%
(\varepsilon_{i}-1)\right)  \log\left(  1+\alpha_{0}(\varepsilon
_{i}-1)\right)  \right]  \right)  ^{-1}. \label{eq def c0}%
\end{equation}

\end{lemma}

\begin{remark}
Note that for the simple case of the ACD of order one where $\psi_{i}%
(\theta)=\omega+\alpha x_{i-1}$, it follows that $c_{0}=\omega_{0}%
/\mathbb{E}\left[  \varepsilon_{1}\log\left(  \varepsilon_{1}\right)  \right]
$ when $\alpha_{0}=1$. Moreover, for exponentially distributed $\varepsilon
_{i}$, ${\mathbb{E}}\left[  \varepsilon_{1}\log\left(  \varepsilon_{1}\right)
\right]  =1-\gamma_{e}$, where $\gamma_{e}\simeq0.577$ is Euler's constant and
hence $c_{0}\simeq2.36\times\omega_{0}$.
\end{remark}

\section{Asymptotic distribution of the QMLE for IACD}

\label{Sec main}

Using the central result in Lemma \ref{lem n div by g} that the number of
observations $n(t)$ is proportional to $t/\log t$ for the integrated ACD
process, we establish in Section \ref{sec unrestricted QMLE} below that the
unrestricted (Q)MLE\ $\hat{\theta}_{t}$ for the ACD\ model is asymptotically
Gaussian when $\alpha_{0}+\beta_{0}=1$. In Section \ref{sec restricted QMLE}
we discuss (Q)ML estimation under the IACD\ restriction $\alpha_{0}+\beta
_{0}=1$ and discuss $t$- and LR-based test statistics for this restriction.
Finally, in Section \ref{ssec testing}\ we revisit these tests and show how to
implement tests of infinite expectation of durations.

\subsection{Asymptotics for the unrestricted QMLE}

\label{sec unrestricted QMLE}

The main result about QMLE of the ACD\ model is given in Theorem
\ref{thm QMLE U} and shows that, while the limiting distribution is standard
(i.e. Gaussian), the rate of convergence $\sqrt{t/\log t},$ is indeed non-standard.

As is common for asymptotic likelihood theory, the asymptotic behavior of the
score and the information determine the limiting results. In terms of the
likelihood function $\mathcal{L}_{n(t)}(\theta)=\sum_{i=1}^{n(t)}\ell
_{i}(\theta)$ in (\ref{eq def Likelihood}), introduce the notation
$\mathcal{S}_{n(t)}=\mathcal{S}_{n(t)}(\theta_{0})$ and $\mathcal{I}%
_{n(t)}=\mathcal{I}_{n(t)}(\theta_{0})$ for the score and information
respectively evaluated at the true parameter $\theta_{0}$, where
\begin{equation}
\mathcal{S}_{n(t)}(\theta)=\sum_{i=1}^{n(t)}s_{i}(\theta)\text{, \ \ and
\ \ \ }\mathcal{I}_{n(t)}(\theta)=\sum_{i=1}^{n(t)}\iota_{i}(\theta)\text{,}
\label{eq score infor}%
\end{equation}
with $s_{i}(\theta)=\partial\ell_{i}(\theta)/\partial\theta$ and $\iota
_{i}(\theta)=-\partial^{2}\ell_{i}(\theta)/\partial\theta\partial
\theta^{\prime}$.

With the ACD\ model given by (\ref{eq ACD model x})-(\ref{eq ACD model int})
we have the following result for the QMLE $\hat{\theta}_{t}$.

\begin{theorem}
\label{thm QMLE U}Consider the QMLE estimator $\hat{\theta}_{t}$ defined in
(\ref{eq def QMLE U}) for the ACD model for $x_{i}>0$ as given by
(\ref{eq ACD model x})-(\ref{eq ACD model int}). With $\theta_{0}=\left(
\omega_{0},\alpha_{0},\beta_{0}\right)  ^{\prime}$ satisfying $\omega
_{0},\alpha_{0}>0$ and the IACD\ hypothesis $\beta_{0}=1-\alpha_{0}>0$, as
$t\rightarrow\infty,$
\begin{equation}
\sqrt{t/\log t}(\hat{\theta}_{t}-\theta_{0})\rightarrow_{d}\operatorname*{N}%
\left(  0,\Sigma\right)  \text{, \ } \label{eq QMLE limit U}%
\end{equation}
where $\Sigma=c_{0}\sigma_{\varepsilon}^{2}\Omega^{-1}$, with $c_{0}$ is
defined in (\ref{eq def c0}) and $\Omega=\mathbb{E}\left[  \iota_{t}%
(\theta_{0})\right]  $, cf. (\ref{eq score infor}).
\end{theorem}

Note in particular, as already emphasized, that while $\hat{\theta}_{t}$
converges at the lower non-standard rate $\sqrt{t/\log t}$, the limiting
distribution is Gaussian.

\begin{remark}
Note that for the case of exponentially distributed innovations,
$\sigma_{\varepsilon}^{2}=1$ and the asymptotic variance simplifies to
$c_{0}\Omega^{-1}$. For the general QMLE\ case, a consistent estimator
$\hat{\Sigma}_{t}$ of\textbf{ }$\Sigma$\textbf{ }in (\ref{eq QMLE limit U}%
)\ is%
\begin{equation}
\hat{\Sigma}_{t}=\tfrac{\log t}{t}\hat{\sigma}_{\varepsilon}^{2}\left[
\mathcal{I}_{n(t)}(\hat{\theta}_{t})\right]  ^{-1}\text{,}
\label{eq cons Sigma}%
\end{equation}
where $\hat{\sigma}_{\varepsilon}^{2}=n\left(  t\right)  ^{-1}\sum
_{i=1}^{n\left(  t\right)  }\left(  \hat{\varepsilon}_{i}-\bar{\varepsilon
}_{n\left(  t\right)  }\right)  ^{2}$, with $\hat{\varepsilon}_{i}=x_{i}%
/\psi_{i}(\hat{\theta}_{t})$ and $\bar{\varepsilon}_{n\left(  t\right)
}=n\left(  t\right)  ^{-1}\sum_{i=1}^{n\left(  t\right)  }\hat{\varepsilon
}_{i}$, is the sample variance of the standardized\ (unrestricted) residuals.
As an alternative to $\hat{\Sigma}_{t}$ as defined in (\ref{eq cons Sigma}),
one may use the well-known asymptotically equivalent estimator
\[
\hat{\Sigma}_{t}=\tfrac{\log t}{t}\left[  \mathcal{I}_{n(t)}(\hat{\theta}%
_{t})\right]  ^{-1}\left[  \sum_{i=1}^{n(t)}s_{i}(\hat{\theta}_{t})s_{i}%
(\hat{\theta}_{t})^{\prime}\right]  \left[  \mathcal{I}_{n(t)}(\hat{\theta
}_{t})\right]  ^{-1}\text{,}%
\]
with $s_{i}(\theta)$ defined in (\ref{eq score infor}).
\end{remark}

\begin{remark}
The results in Theorem \ref{thm QMLE U} also applies to estimation of the
simple ACD model of order one where $\psi_{i}\left(  \theta\right)
=\omega+\alpha x_{i-1}$. That is, with $\theta=\left(  \omega,\alpha\right)
^{\prime}$, and $\hat{\theta}_{t}$ the QMLE for the order one ACD, then
$\sqrt{t/\log t}(\hat{\theta}_{t}-\theta_{0})$ converges as in
(\ref{eq QMLE limit U}).
\end{remark}

Theorem \ref{thm QMLE U} is stated in terms of the deterministic normalization
$(\log t)/t$. Alternatively, one may state convergence results in terms of
either of the following two random statistics:\ the number of observations,
$n(t)$, or the information, $\mathcal{I}_{n(t)}$. That is, as an immediate
implication of Theorem \ref{thm QMLE U} we have the following corollary.

\begin{corollary}
\label{cor random norrm QMLE}Consider the QMLE estimator $\hat{\theta}_{t}$
defined in (\ref{eq def QMLE U}). Under the assumptions of Theorem
\ref{thm QMLE U}, as $t\rightarrow\infty$, $\sqrt{n(t)}(\hat{\theta}%
_{t}-\theta_{0})\rightarrow_{d}\operatorname*{N}\left(  0,\sigma_{\varepsilon
}^{2}\Omega^{-1}\right)  $, and $\mathcal{I}_{n(t)}^{1/2}(\hat{\theta}%
_{t}-\theta_{0})\rightarrow_{d}\operatorname*{N}\left(  0,\sigma_{\varepsilon
}^{2}I_{3}\right)  $, with $I_{3}$ denoting the $\left(  3\times3\right)
$-dimensional identity matrix.
\end{corollary}

The latter result in Corollary \ref{cor random norrm QMLE} means in particular
that for the MLE, where $\sigma_{\varepsilon}^{2}=1$, $\mathcal{I}%
_{n(t)}^{1/2}(\hat{\theta}_{t}-\theta_{0})$ is asymptotically standard
Gaussian distributed.

A further immediate implication of the results in Theorem \ref{thm QMLE U} is
that we can state the limiting distribution of the $t$-statistic for testing
the hypothesis of integrated ACD.

\begin{corollary}
\label{cor tstat QMLE}Consider the $t$-statistic as defined by
\begin{equation}
\tau_{t}=\sqrt{t/\log t}\frac{\hat{\alpha}_{t}+\hat{\beta}_{t}-1}{(g^{\prime
}\hat{\Sigma}_{t}g)^{1/2}} \label{eq def t stat}%
\end{equation}
where $g=\left(  0,1,1\right)  ^{\prime}$ and $\hat{\Sigma}_{t}$ is defined in
(\ref{eq cons Sigma}). Under the assumptions of Theorem \ref{thm QMLE U}, as
$t\rightarrow\infty$, $\tau_{t}\rightarrow_{d}\operatorname*{N}\left(
0,1\right)  $.
\end{corollary}

\subsection{Asymptotics for the restricted QMLE}

\label{sec restricted QMLE}

Next, turn to QML estimation under the restriction $\alpha+\beta=1$,
corresponding to the IACD\ model. Introduce the parameter vector $\phi
\in\mathbb{R}^{2}$, where $\phi=\left(  \phi_{1},\phi_{2}\right)  ^{\prime
}=\left(  \omega,\alpha\right)  ^{\prime}\in\Phi\subset\lbrack0,\infty)^{2}$,
such that $\theta=\theta\left(  \phi\right)  =\left(  \omega,\alpha
,1-\alpha\right)  ^{\prime}\in\Theta$ for $\phi\in\Phi$. The restricted QML
estimator $\tilde{\theta}_{t}=(\Tilde{\omega}_{t},\tilde{\alpha}_{t}%
,\tilde{\beta}_{t})^{\prime}\ $is then given by $\tilde{\theta}_{t}%
=\theta(\tilde{\phi}_{t})$, where
\[
\tilde{\phi}_{t}=\arg\max_{\phi\in\Phi}\mathcal{L}_{n(t)}\left(  \theta\left(
\phi\right)  \right)  \text{,}%
\]
with $\mathcal{L}_{n(t)}(\theta)$ defined in (\ref{eq def QMLE U}). For this
estimator, we have the following result.

\begin{theorem}
\label{thm QMLE R and QLR}Under the assumptions of Theorem \ref{thm QMLE U},
with $\phi=\left(  \phi_{1},\phi_{2}\right)  ^{\prime}=\left(  \omega
,\alpha\right)  ^{\prime}$, $\omega_{0}>0$,\textbf{ }and $0<\alpha_{0}<1$, it
follows that for the QMLE $\tilde{\phi}_{t}$ of $\phi$, as $t\rightarrow
\infty$,%
\begin{equation}
\sqrt{t/\log t}(\tilde{\phi}_{t}-\phi_{0})\rightarrow_{d}\operatorname*{N}%
(0,\Sigma_{\phi})\text{,} \label{eq R QMLE}%
\end{equation}
where $\Sigma_{\phi}=c_{0}\sigma_{\varepsilon}^{2}\left(  \gamma^{\prime
}\Omega\gamma\right)  ^{-1},\gamma=\partial\theta\left(  \phi\right)
/\partial\phi^{\prime}$ is given by (\ref{def gamma}),\ and $c_{0}$ is given
by (\ref{eq def c0}). Moreover, the quasi-likelihood ratio statistic
$\operatorname*{QLR}_{t}$ satisfies, as $t\rightarrow\infty$,
\begin{equation}
\operatorname*{QLR}\nolimits_{t}=2\left[  \mathcal{L}_{n(t)}(\hat{\theta}%
_{t})-\mathcal{L}_{n(t)}(\theta(\Tilde{\phi}_{t}))\right]  \rightarrow
_{d}\sigma_{\varepsilon}^{2}\chi_{1}^{2}\text{.} \label{def QLR stat}%
\end{equation}

\end{theorem}

\begin{remark}
In line with the $\operatorname*{QLR}_{t}$ statistic for the hypothesis of
integrated ACD in Theorem \ref{thm QMLE R and QLR}, we note that the analogous
statistic is considered by simulations for the GARCH model in Busch~(2005) and
Lumsdaine~(1995). Moreover, whereas restricted estimation is to our knowledge
not covered in existing literature, unrestricted estimation theory which
allows integrated GARCH, is considered in Berkes, Horvath and Kokoszka~(2003),
Lee and Hansen~(1994) and Lumsdaine~(1996). We emphasize that the theory in
the mentioned papers does not apply to the case of ACD\ models due to the
random number of observations $n(t)$.
\end{remark}

\subsection{Testing IACD and infinite expectation of durations}

\label{ssec testing}

Consider initially testing the null hypothesis of IACD, i.e., $\mathsf{H}%
_{\text{IACD}}:\alpha+\beta=1$, against the alternative $\alpha+\beta\neq1$.
To this aim, it is natural to use the $\operatorname*{QLR}_{t}$ statistic in
(\ref{def QLR stat}), which, normalized by $\hat{\sigma}_{\varepsilon}^{2}$,
is asymptotically $\chi_{1}^{2}$-distributed. Alternatively, one may run a
two-sided test based on the $\tau_{t}$ statistic in (\ref{eq def t stat}),
which is asymptotically standard normal under the null.

As an alternative, one may do one-sided testing based on $\tau_{t}$, similar
to the tests for finite moments in GARCH models discussed in Francq and
Zakoian~(2022). Specifically, consider testing the null of infinite
expectation, that is, $\mathsf{H}_{\infty}$: $\alpha+\beta\geq1$ (or
$\mathbb{E}\left[  x_{i}\right]  =\infty$) against the alternative
$\alpha+\beta<1$ (or $\mathbb{E}\left[  x_{i}\right]  <\infty)$. In this case,
at nominal level $\eta$, with critical value $q\left(  \eta\right)  =\Phi
^{-1}\left(  \eta\right)  $, $\Phi\left(  \cdot\right)  $ being the standard
normal distribution function, the null $\mathsf{H}_{\infty}$ is rejected
provided $\tau_{t}<q\left(  \eta\right)  $. This implies that the asymptotic
size of the test is less than or equal to $\eta$, and in this sense size is
controlled as $t\rightarrow\infty$.

Although of less interest here, one may also test the null of $\alpha
+\beta\leq1$ against the alternative of $\alpha+\beta>1$, in which case one
rejects when $\tau_{t}>q\left(  1-\eta\right)  $. This is of less interest as
there is no direct interpretation in terms of (in)finite expectation; in
particular, it could be viewed as a test for the null of finite expectation,
$\mathbb{E}[x_{i}]<\infty$, but such a test would have power only against
alternatives with $\alpha_{0}+\beta_{0}>1$, while having power equal to size
in the IACD\ case $\alpha_{0}+\beta_{0}=1$. As for the previous test, also for
this test the asymptotic size is less than or equal to the nominal level
$\eta$.

These different testing scenarios are investigated using Monte Carlo
simulation in Section \ref{sec MC}, and applied to cryptocurrency data in the
Section \ref{sec empirical}.

\section{Simulations}
\label{sec MC}

Using Monte Carlo simulation, in this section we assess the finite sample
performance and accuracy of the asymptotic properties of testing based on the
$\tau_{t}$ and $\operatorname{QLR}_{t}$ statistics as discussed in the
previous section. In particular, we want to analyze both the finite-sample
behavior of these statistics under the IACD null hypothesis and their behavior
under alternative hypotheses featuring both finite and infinite expected durations.

In Section \ref{sec MC set up} we introduce the Monte Carlo set up, including
computational details for the reference tests. In Section \ref{sec MC size} we
present the behavior of the test statistics under the null hypothesis (size),
while in Section \ref{sec MC power} we discuss results under the alternative (power).

\subsection{Set up}

\label{sec MC set up}

The data generating processes (DGPs) for the simulations are given by the ACD
process as defined in equations (\ref{eq ACD model x})-(\ref{eq ACD model int}%
), with the intercept parameter fixed at $\omega_{0}=1$. We vary the
parameters $\alpha_{0}$ and $\beta_{0}$ in the strict stationarity region,
such that either the key IACD condition $\alpha_{0}+\beta_{0}=1$ holds, or the
parameters satisfy the inequality conditions $\alpha_{0}+\beta_{0}>1$ (such
that $\mathbb{E}[x_{i}]=+\infty$ and the tail index $\kappa_{0}$ is below
unity) or $\alpha_{0}+\beta_{0}<1$ (such that $\mathbb{E}[x_{i}]<+\infty$ and
the tail index $\kappa_{0}$ is above unity). Thus, we consider scenarios
reflecting both IACD and non-integrated ACD processes to evaluate the behavior
of the test statistics under different tail indices.

Specifically, for a given time span $[0,t]$, $t>0$, durations $\left\{
x_{i}\right\}  _{i=0}^{n(t)}$ are simulated using a `burn-in' sample of
$b=1000$ observations. That is, the $x_{i}$'s are generated recursively as%
\begin{align*}
x_{i}  &  =\psi_{i}\varepsilon_{i}\text{, \ \ }i=-(b-1),\ldots,-1,0,1,\ldots
,n\left(  t\right) \\
\psi_{i}  &  =\omega_{0}+\alpha_{0}x_{i-1}+\beta_{0}\psi_{i-1},
\end{align*}
with $\left\{  \varepsilon_{i}\right\}  $ is an i.i.d. sequence of strictly
positive random variables with $\mathbb{E}[\varepsilon_{i}]=1$ and initial
values $x_{-b}=\psi_{-b}=0$.\textbf{ }Estimators and test statistics are based
on the\textbf{ }likelihood function in (\ref{eq def Likelihood}). The
observations entering the likelihood function are the simulated\textbf{
}$\left\{  x_{i}\right\}  _{i=1}^{n\left(  t\right)  }$\textbf{ }with, as also
done in the empirical illustration in Section \ref{sec empirical}, initial
values\textbf{ }$\left(  x_{0},\psi_{0}\left(  \theta\right)  \right)
=\left(  x_{0},x_{0}\right)  $.\textbf{ }One may alternatively use $\psi
_{0}\left(  \theta\right)  =\omega$ in the likelihood function (see, e.g.,
Francq and Zako\"{\i}an, 2019, Ch.7); however, we found no discernible
differences in applying either of the two choices.

In order to evaluate the impact of the shape of the distribution of the
innovations $\varepsilon_{i}$, we consider Weibull distributed random
variables with shape parameter $\nu>0$, scaled such that $\mathbb{E}%
[\varepsilon_{i}]=1$. The associated probability density function (pdf) is
given by
\[
f_{\varepsilon,v}(x)=\nu\Gamma(1+1/\nu)^{\nu}x^{\nu-1}\exp(-(x\Gamma
(1+1/\nu))^{\nu}),\text{ \ \ \ }x\geq0,
\]
where $\Gamma\left(  \cdot\right)  $ is the Gamma function.\ Note that
$f_{\varepsilon,\nu}\left(  \cdot\right)  $ reduces to the pdf of the standard
exponential distribution for $\nu=1$. Moreover, the variance of $\varepsilon
_{i}$ as a function of $\nu$, $\sigma_{\varepsilon}^{2}(\nu)$, is given by%
\[
\sigma_{\varepsilon}^{2}(\nu)=\tfrac{\Gamma(1+2/\nu)}{\Gamma(1+1/\nu)^{2}%
}-1\text{,}%
\]
implying that $\sigma_{\varepsilon}^{2}(\nu)$ decreases monotonically with
respect to $\nu$ and achieves the value $\sigma_{\varepsilon}^{2}(1)=1$ (the
exponential distribution). Moreover, $\lim_{\nu\rightarrow\infty}%
\sigma_{\varepsilon}^{2}(\nu)=0$ and $\lim_{\nu\rightarrow0}\sigma
_{\varepsilon}^{2}(\nu)=\infty$. By varying $\nu$, we include under-dispersion
for $\nu>1$ ($\sigma_{\varepsilon}^{2}\left(  v\right)  <1$), the exponential
case of $\nu=1$, and over-dispersion for $\nu<1$ ($\sigma_{\varepsilon}%
^{2}\left(  v\right)  >1$). For all designs, the number of Monte Carlo
replications is $M=10,000$ and the nominal level of tests is $\eta=0.05$, cf.
Section \ref{ssec testing}.

\subsection{Properties of tests under the null}

\label{sec MC size}

In line with the discussion in Section \ref{ssec testing}, we consider first
results for two-sided tests for the IACD\ null hypothesis, based on the
statistics $\tau_{t}$ and $\operatorname*{QLR}_{t}$. Later, we turn to
one-sided tests based on $\tau_{t}$.

We consider parameter settings where $\alpha_{0}\in\{0.15,0.50,0.85\}$ and
$\beta_{0}=1-\alpha_{0}$. The shape parameter $\nu$ of the Weibull
distribution takes values in the set $\{1.435,1.000,0.721\}$ such that
$\sigma_{\varepsilon}^{2}\left(  \nu\right)  \in\{0.5,1.0,2.0\}$, representing
under- and over-dispersion, as well as the exponential case. For each
combination of $\left(  v,\alpha_{0}\right)  $, we consider five different
values of the time span length $t$, selected by calibrating the
(simulated)\ median number of events $\operatorname*{med}\left\{  n\left(
t\right)  \right\}  $ to be in the set $\{100,500,2500,12500,62500\}$. The
latter two are close to the number of durations in the empirical illustration.

\subsubsection{Testing IACD}%

\begin{table}[t] \centering
\caption{\textsc{Empirical Rejection Probabilities under the Null Hypothesis -- two-sided tests.}}
\vspace{0.5em}%
\begin{tabular}
[c]{cccccccc}\hline
\multicolumn{2}{c}{} & \multicolumn{2}{c}{$\alpha_{0}=0.15$} &
\multicolumn{2}{c}{$\alpha_{0}=0.50$} & \multicolumn{2}{c}{$\alpha_{0}=0.85$%
}\\\hline
$\sigma_{\varepsilon}^{2}\left(  \nu\right)  $ & $\operatorname*{med}\left\{
n\left(  t\right)  \right\}  $ & $\tau_{t}$ & $\operatorname*{QLR}_{t}$ &
$\tau_{t}$ & $\operatorname*{QLR}_{t}$ & $\tau_{t}$ & $\operatorname*{QLR}%
_{t}$\\\hline
\multicolumn{1}{r}{$\overset{}{0.5}$} & \multicolumn{1}{r}{$100$} &
\multicolumn{1}{r}{$0.226$} & \multicolumn{1}{r}{$0.378$} &
\multicolumn{1}{r}{$0.079$} & \multicolumn{1}{r}{$0.121$} &
\multicolumn{1}{r}{$0.060$} & \multicolumn{1}{r}{$0.070$}\\
\multicolumn{1}{r}{} & \multicolumn{1}{r}{$500$} & \multicolumn{1}{r}{$0.160$}
& \multicolumn{1}{r}{$0.196$} & \multicolumn{1}{r}{$0.048$} &
\multicolumn{1}{r}{$0.049$} & \multicolumn{1}{r}{$0.046$} &
\multicolumn{1}{r}{$0.056$}\\
\multicolumn{1}{r}{} & \multicolumn{1}{r}{$2500$} & \multicolumn{1}{r}{$0.067$%
} & \multicolumn{1}{r}{$0.069$} & \multicolumn{1}{r}{$0.054$} &
\multicolumn{1}{r}{$0.059$} & \multicolumn{1}{r}{$0.058$} &
\multicolumn{1}{r}{$0.061$}\\
\multicolumn{1}{r}{} & \multicolumn{1}{r}{$12500$} &
\multicolumn{1}{r}{$0.053$} & \multicolumn{1}{r}{$0.053$} &
\multicolumn{1}{r}{$0.056$} & \multicolumn{1}{r}{$0.059$} &
\multicolumn{1}{r}{$0.057$} & \multicolumn{1}{r}{$0.059$}\\
& \multicolumn{1}{r}{$62500$} & $0.063$ & $0.065$ & $0.054$ & $0.056$ &
$0.051$ & $0.053$\\
\multicolumn{1}{r}{$\overset{}{1.0}$} & \multicolumn{1}{r}{$100$} &
\multicolumn{1}{r}{$0.145$} & \multicolumn{1}{r}{$0.340$} &
\multicolumn{1}{r}{$0.063$} & \multicolumn{1}{r}{$0.103$} &
\multicolumn{1}{r}{$0.056$} & \multicolumn{1}{r}{$0.070$}\\
\multicolumn{1}{r}{} & \multicolumn{1}{r}{$500$} & \multicolumn{1}{r}{$0.130$}
& \multicolumn{1}{r}{$0.165$} & \multicolumn{1}{r}{$0.044$} &
\multicolumn{1}{r}{$0.052$} & \multicolumn{1}{r}{$0.046$} &
\multicolumn{1}{r}{$0.059$}\\
\multicolumn{1}{r}{} & \multicolumn{1}{r}{$2500$} & \multicolumn{1}{r}{$0.061$%
} & \multicolumn{1}{r}{$0.064$} & \multicolumn{1}{r}{$0.051$} &
\multicolumn{1}{r}{$0.059$} & \multicolumn{1}{r}{$0.056$} &
\multicolumn{1}{r}{$0.061$}\\
\multicolumn{1}{r}{} & \multicolumn{1}{r}{$12500$} &
\multicolumn{1}{r}{$0.059$} & \multicolumn{1}{r}{$0.062$} &
\multicolumn{1}{r}{$0.057$} & \multicolumn{1}{r}{$0.060$} &
\multicolumn{1}{r}{$0.056$} & \multicolumn{1}{r}{$0.058$}\\
& \multicolumn{1}{r}{$62500$} & $0.060$ & $0.063$ & $0.054$ & $0.056$ &
$0.050$ & $0.052$\\
\multicolumn{1}{r}{$\overset{}{2.0}$} & \multicolumn{1}{r}{$100$} &
\multicolumn{1}{r}{$0.082$} & \multicolumn{1}{r}{$0.294$} &
\multicolumn{1}{r}{$0.054$} & \multicolumn{1}{r}{$0.097$} &
\multicolumn{1}{r}{$0.053$} & \multicolumn{1}{r}{$0.075$}\\
\multicolumn{1}{r}{} & \multicolumn{1}{r}{$500$} & \multicolumn{1}{r}{$0.094$}
& \multicolumn{1}{r}{$0.131$} & \multicolumn{1}{r}{$0.045$} &
\multicolumn{1}{r}{$0.056$} & \multicolumn{1}{r}{$0.049$} &
\multicolumn{1}{r}{$0.060$}\\
\multicolumn{1}{r}{} & \multicolumn{1}{r}{$2500$} & \multicolumn{1}{r}{$0.057$%
} & \multicolumn{1}{r}{$0.064$} & \multicolumn{1}{r}{$0.048$} &
\multicolumn{1}{r}{$0.058$} & \multicolumn{1}{r}{$0.052$} &
\multicolumn{1}{r}{$0.061$}\\
\multicolumn{1}{r}{} & \multicolumn{1}{r}{$12500$} &
\multicolumn{1}{r}{$0.060$} & \multicolumn{1}{r}{$0.066$} &
\multicolumn{1}{r}{$0.055$} & \multicolumn{1}{r}{$0.059$} &
\multicolumn{1}{r}{$0.050$} & \multicolumn{1}{r}{$0.052$}\\
& \multicolumn{1}{r}{$62500$} & $0.057$ & $0.059$ & $0.054$ & $0.056$ &
$0.050$ & $0.052$\\\hline
\end{tabular}
\ \vspace{-0.5em}

\begin{flushleft}
{\small \textit{Notes:} The table reports empirical rejection probabilities
for the $\operatorname*{QLR}_{t}$ and $\tau_{t}$ statistics under the null
hypothesis $\alpha_{0} + \beta_{0} = 1$ (IACD). Results are based on $M =
10,000$ Monte Carlo replications. }
\end{flushleft}

\label{table RP under null}%
\end{table}%

In Table \ref{table RP under null} we report the (simulated) empirical
rejection probabilities (ERPs) under the null hypothesis of IACD,
$\mathsf{H}_{\text{IACD}}$, using both statistics ($\tau_{t}$ and
$\operatorname*{QLR}_{t}$). We observe that for moderate to large sample sizes
($\operatorname{med}\left\{  n(t)\right\}  \geq2500$), both the $\tau_{t}$ and
$\operatorname{QLR}_{t}$ statistics show rejection frequencies close to the
nominal level across all values of $\left(  \alpha_{0},\beta_{0}\right)  $ and
of the shape parameter $\nu$. Some size distortions are present for shorter
samples, especially in the case of over-dispersion ($\sigma_{\varepsilon}%
^{2}\left(  v\right)  =2>1$) and for stronger persistence (e.g., $\alpha
_{0}=0.15$), where the $\operatorname*{QLR}_{t}$-based test is oversized, in
particular relatively to $\tau_{t}$. Nonetheless, both tests demonstrate good
finite-sample size control in all reasonable settings.

\begin{figure}[t]
\begin{center}
\includegraphics[
trim=0.000000in 0.005504in 0.002033in -0.005504in,
height=8.0cm,
width=12.0cm
]{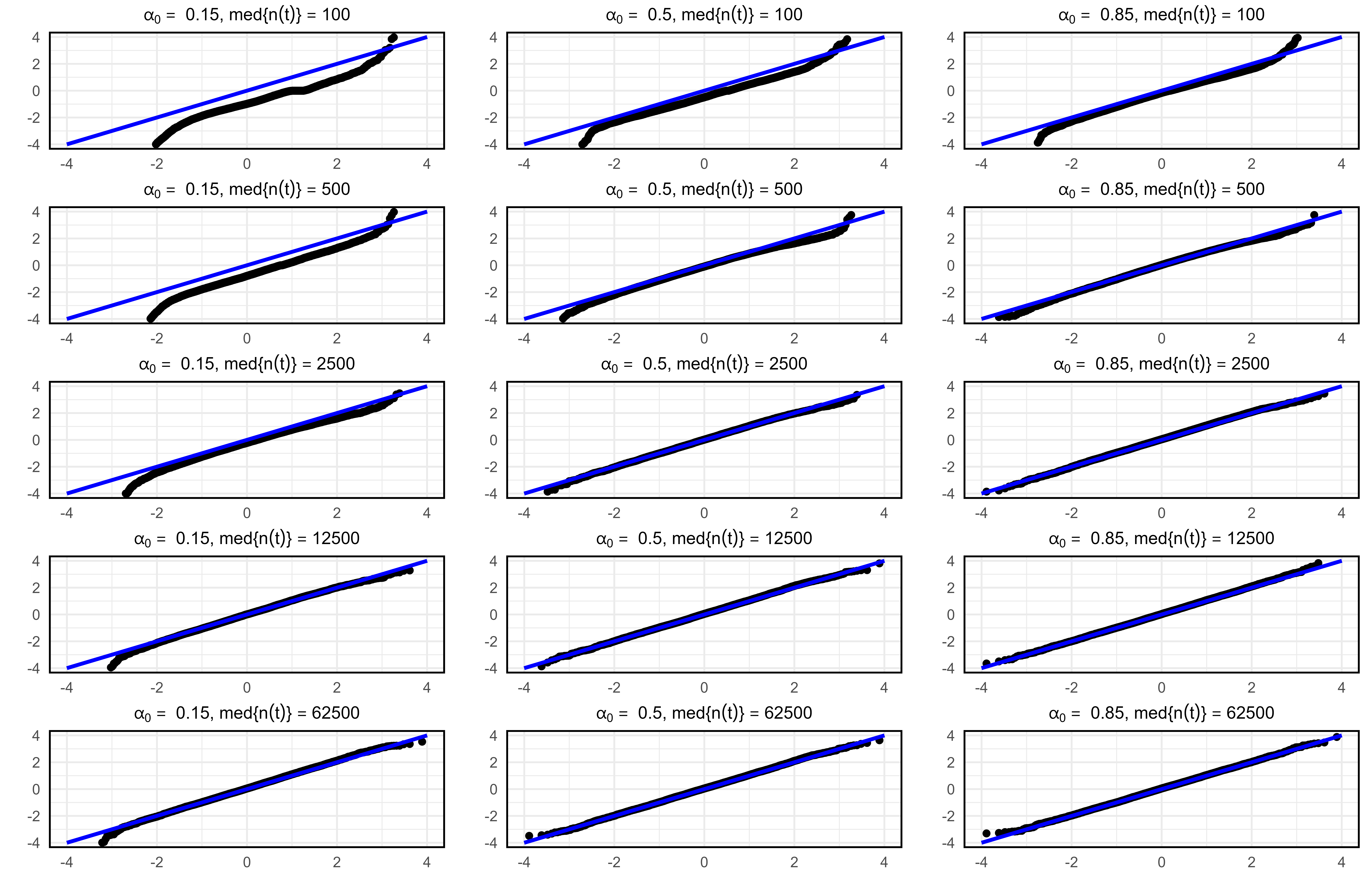}
\end{center}
\caption{\textsc{QQ-plot.} Quantiles of the $\tau_{t}$ statistic in
(\ref{eq def t stat}) against the $N(0,1)$-distribution. For each $\alpha
_{0}\in\{0.15,0.5,0.85\}$, values of $t$ are such that median number of
observations, $\operatorname*{med}\left\{  n\left(  t\right)  \right\}  \in$
$\left\{  100,500,2500,12500,62500\right\}  $. Simulations based on
$\varepsilon_{i}$ exponentially distributed. Number of Monte
Carlo-replications $M=10000.$}%
\label{fig QQ plot2}%
\end{figure}

To further illustrate the validity of the asymptotic results, we consider the
$\tau_{t}$ statistic which by Corollary \ref{cor tstat QMLE} is asymptotically
standard normal. Figure \ref{fig QQ plot2} provides QQ-plots of the $\tau_{t}$
statistic against the standard normal distribution for the case of $\nu=1$ and
$\alpha_{0}\in\left\{  0.15,0.5,0.85\right\}  $. The QQ-plots confirm the
findings reported in Table \ref{table RP under null}. In particular, the
quality of the standard Gaussian approximation improves markedly as the median
number of observations increases. For small sample sizes (e.g.,
$\operatorname{med}\left\{  n(t)\right\}  =100$), the quantiles of $\tau_{t}$
deviate substantially from the standard normal, particularly in the tails and
for $\alpha_{0}=0.15$. As the median increases to $2500$ and beyond, the
empirical quantiles align much more closely with the Gaussian quantiles, as
predicted by the asymptotic theory. This pattern holds across all values of
$\alpha_{0}$, though convergence is slower for $\alpha_{0}=0.15$. Overall, the
figure confirms our theoretical finds, also highlighting the importance of
sufficiently large sample sizes for reliable inference in practice, probably
due to the slower convergence rate of estimators when $\alpha_{0}+\beta
_{0}\geq1$.

\subsubsection{One-sided testing}

In Table \ref{table RP under null one sided} we report the ERPs for the
one-sided tests. Here, with $q=q\left(  0.95\right)  \simeq1.64$, the columns
$\tau_{t}<-q$ report the ERPs for the case of testing $\mathsf{H}_{\infty
}:\alpha+\beta\geq1$ against the alternative $\alpha+\beta<1$, while the
columns with $\tau_{t}>q$, report the ERPs for testing $\alpha+\beta\leq1$
against the alternative $\alpha+\beta>1$. As for the two-sided tests, the size
appears well controlled for sufficiently large samples. A noticeable
difference between the two tests is that the (left sided) test for the
infinite expected duration hypothesis, $\mathsf{H}_{\infty}:\alpha+\beta\geq
1$, the ERPs are above the nominal level for the smaller sample sizes, while
the right-sided test is undersized in small samples. For both tests, the ERPs
tend to the nominal level $\eta=0.05$ as the median number of observations,
$\operatorname*{med}\left\{  n\left(  t\right)  \right\}  $, increases.%

\begin{table}[t] \centering
\caption{\textsc{Empirical Rejection Probabilities under the Null Hypothesis -- one-sided tests.}}
\vspace{0.5em}%
\begin{tabular}
[c]{cccccccc}\hline
\multicolumn{2}{c}{} & \multicolumn{2}{c}{$\alpha_{0}=0.15$} &
\multicolumn{2}{c}{$\alpha_{0}=0.50$} & \multicolumn{2}{c}{$\alpha_{0}=0.85$%
}\\\hline
$\sigma_{\varepsilon}^{2}\left(  \nu\right)  $ & $\operatorname*{med}\left\{
n\left(  t\right)  \right\}  $ & $\tau_{t}<-q$ & $\tau_{t}>q$ & $\tau_{t}<-q$
& $\tau_{t}>q$ & $\tau_{t}<-q$ & $\tau_{t}>q$\\\hline
\multicolumn{1}{r}{$\overset{}{0.5}$} & \multicolumn{1}{r}{$100$} &
\multicolumn{1}{r}{$0.304$} & \multicolumn{1}{r}{$0.006$} &
\multicolumn{1}{r}{$0.131$} & \multicolumn{1}{r}{$0.016$} &
\multicolumn{1}{r}{$0.082$} & \multicolumn{1}{r}{$0.024$}\\
\multicolumn{1}{r}{} & \multicolumn{1}{r}{$500$} & \multicolumn{1}{r}{$0.239$}
& \multicolumn{1}{r}{$0.010$} & \multicolumn{1}{r}{$0.073$} &
\multicolumn{1}{r}{$0.025$} & \multicolumn{1}{r}{$0.056$} &
\multicolumn{1}{r}{$0.045$}\\
\multicolumn{1}{r}{} & \multicolumn{1}{r}{$2500$} & \multicolumn{1}{r}{$0.099$%
} & \multicolumn{1}{r}{$0.018$} & \multicolumn{1}{r}{$0.054$} &
\multicolumn{1}{r}{$0.055$} & \multicolumn{1}{r}{$0.048$} &
\multicolumn{1}{r}{$0.061$}\\
\multicolumn{1}{r}{} & \multicolumn{1}{r}{$12500$} &
\multicolumn{1}{r}{$0.054$} & \multicolumn{1}{r}{$0.050$} &
\multicolumn{1}{r}{$0.050$} & \multicolumn{1}{r}{$0.064$} &
\multicolumn{1}{r}{$0.047$} & \multicolumn{1}{r}{$0.062$}\\
& \multicolumn{1}{r}{$62500$} & \multicolumn{1}{r}{$0.051$} & $0.062$ &
\multicolumn{1}{r}{$0.046$} & \multicolumn{1}{r}{$0.060$} &
\multicolumn{1}{r}{$0.045$} & \multicolumn{1}{r}{$0.053$}\\
\multicolumn{1}{r}{$\overset{}{1.0}$} & \multicolumn{1}{r}{$100$} &
\multicolumn{1}{r}{$0.221$} & \multicolumn{1}{r}{$0.005$} &
\multicolumn{1}{r}{$0.108$} & \multicolumn{1}{r}{$0.013$} &
\multicolumn{1}{r}{$0.079$} & \multicolumn{1}{r}{$0.022$}\\
\multicolumn{1}{r}{} & \multicolumn{1}{r}{$500$} & \multicolumn{1}{r}{$0.196$}
& \multicolumn{1}{r}{$0.010$} & \multicolumn{1}{r}{$0.067$} &
\multicolumn{1}{r}{$0.022$} & \multicolumn{1}{r}{$0.058$} &
\multicolumn{1}{r}{$0.040$}\\
\multicolumn{1}{r}{} & \multicolumn{1}{r}{$2500$} & \multicolumn{1}{r}{$0.091$%
} & \multicolumn{1}{r}{$0.020$} & \multicolumn{1}{r}{$0.053$} &
\multicolumn{1}{r}{$0.053$} & \multicolumn{1}{r}{$0.049$} &
\multicolumn{1}{r}{$0.059$}\\
\multicolumn{1}{r}{} & \multicolumn{1}{r}{$12500$} &
\multicolumn{1}{r}{$0.058$} & \multicolumn{1}{r}{$0.053$} &
\multicolumn{1}{r}{$0.048$} & \multicolumn{1}{r}{$0.060$} &
\multicolumn{1}{r}{$0.049$} & \multicolumn{1}{r}{$0.057$}\\
& \multicolumn{1}{r}{$62500$} & \multicolumn{1}{r}{$0.050$} &
\multicolumn{1}{r}{$0.062$} & \multicolumn{1}{r}{$0.047$} &
\multicolumn{1}{r}{$0.055$} & \multicolumn{1}{r}{$0.046$} &
\multicolumn{1}{r}{$0.052$}\\
\multicolumn{1}{r}{$\overset{}{2.0}$} & \multicolumn{1}{r}{$100$} &
\multicolumn{1}{r}{$0.143$} & \multicolumn{1}{r}{$0.006$} &
\multicolumn{1}{r}{$0.096$} & \multicolumn{1}{r}{$0.012$} &
\multicolumn{1}{r}{$0.079$} & \multicolumn{1}{r}{$0.018$}\\
\multicolumn{1}{r}{} & \multicolumn{1}{r}{$500$} & \multicolumn{1}{r}{$0.151$}
& \multicolumn{1}{r}{$0.008$} & \multicolumn{1}{r}{$0.069$} &
\multicolumn{1}{r}{$0.020$} & \multicolumn{1}{r}{$0.071$} &
\multicolumn{1}{r}{$0.030$}\\
\multicolumn{1}{r}{} & \multicolumn{1}{r}{$2500$} & \multicolumn{1}{r}{$0.088$%
} & \multicolumn{1}{r}{$0.023$} & \multicolumn{1}{r}{$0.056$} &
\multicolumn{1}{r}{$0.046$} & \multicolumn{1}{r}{$0.055$} &
\multicolumn{1}{r}{$0.053$}\\
\multicolumn{1}{r}{} & \multicolumn{1}{r}{$12500$} &
\multicolumn{1}{r}{$0.064$} & \multicolumn{1}{r}{$0.054$} &
\multicolumn{1}{r}{$0.048$} & \multicolumn{1}{r}{$0.055$} &
\multicolumn{1}{r}{$0.051$} & \multicolumn{1}{r}{$0.053$}\\
& \multicolumn{1}{r}{$62500$} & \multicolumn{1}{r}{$0.052$} &
\multicolumn{1}{r}{$0.058$} & \multicolumn{1}{r}{$0.049$} &
\multicolumn{1}{r}{$0.056$} & \multicolumn{1}{r}{$0.049$} &
\multicolumn{1}{r}{$0.054$}\\\hline
\end{tabular}
$\ $\bigskip\ \vspace{-1.5em}

\begin{flushleft}
{\small \textit{Notes:} The table reports empirical rejection probabilities
for the one-sided t-tests based on $\tau_{t}$, with $q=1.64$. See also Table
1. }
\end{flushleft}

\label{table RP under null one sided}%
\end{table}%

\begin{figure}[h]
\begin{center}
\includegraphics[
trim=0.000000in 0.005504in 0.002033in -0.005504in,
height=8.0cm,
width=12.0cm
]{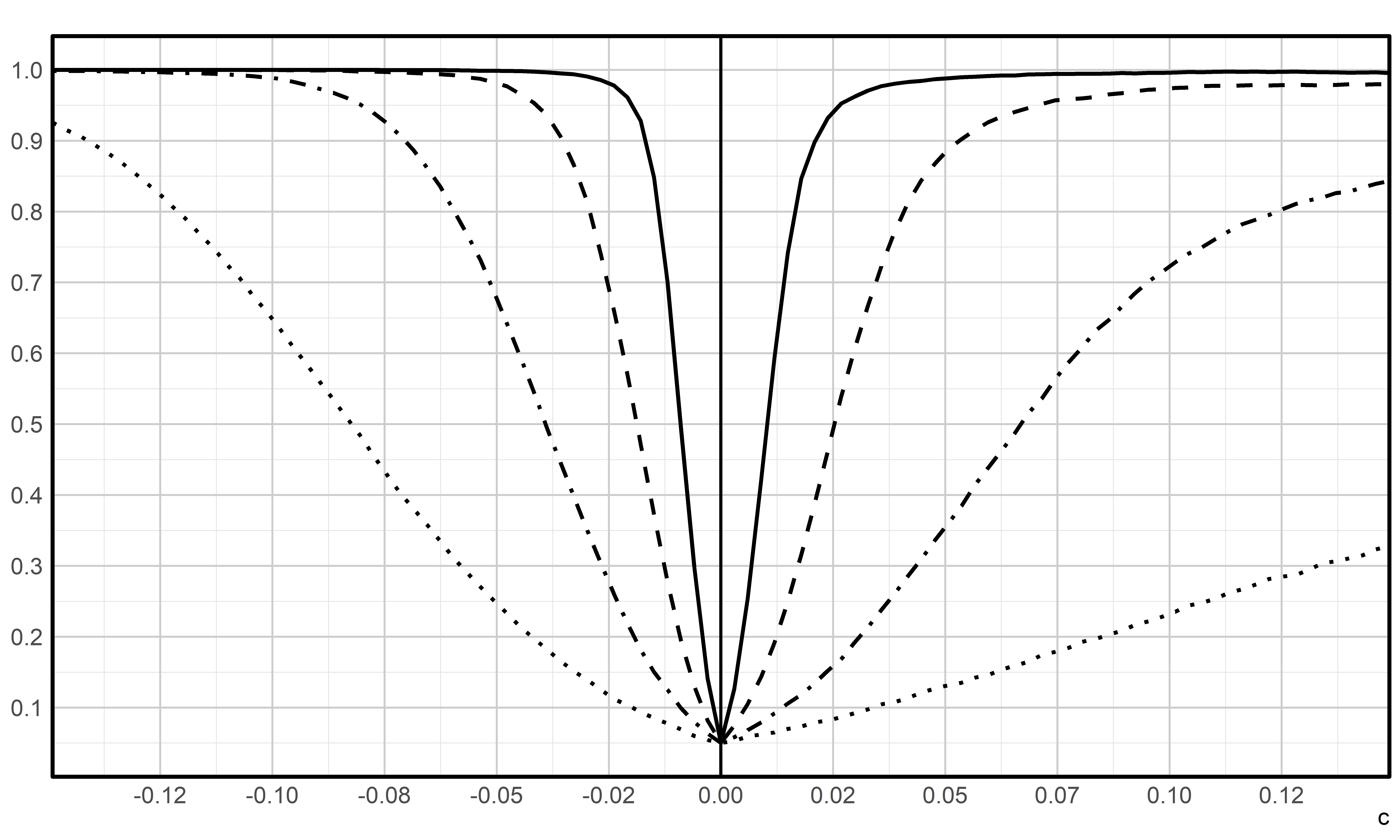}
\end{center}
\caption{\textsc{Rejection frequencies under alternative. }Case of $\alpha
_{0}=0.85$.\ Rejection frequencies for $\tau_{t}>q_{R}$ (right hand side of
$c=0$)\ and $\tau_{t}<q_{L}$(left hand side), with $q_{R}$ [$q_{L}$] the
size-adjusted $0.95$ [$0.05$] quantiles. Solid line: median number of
observations $\operatorname*{med}\{n(t)\}=62500$ for $c=0$; dashed,
dotted-dashed and dotted lines: $\operatorname*{med}\{n(t)\}$ equal to
$12500$, $2500$ and $500$, respectively. Number of Monte Carlo-replications
$M=10000.$}%
\label{RP1}%
\end{figure}

\bigskip

\begin{figure}[ptb]
\begin{center}
\includegraphics[
trim=0.000000in 0.005504in 0.002033in -0.005504in,
height=8.0cm,
width=12.0cm
]{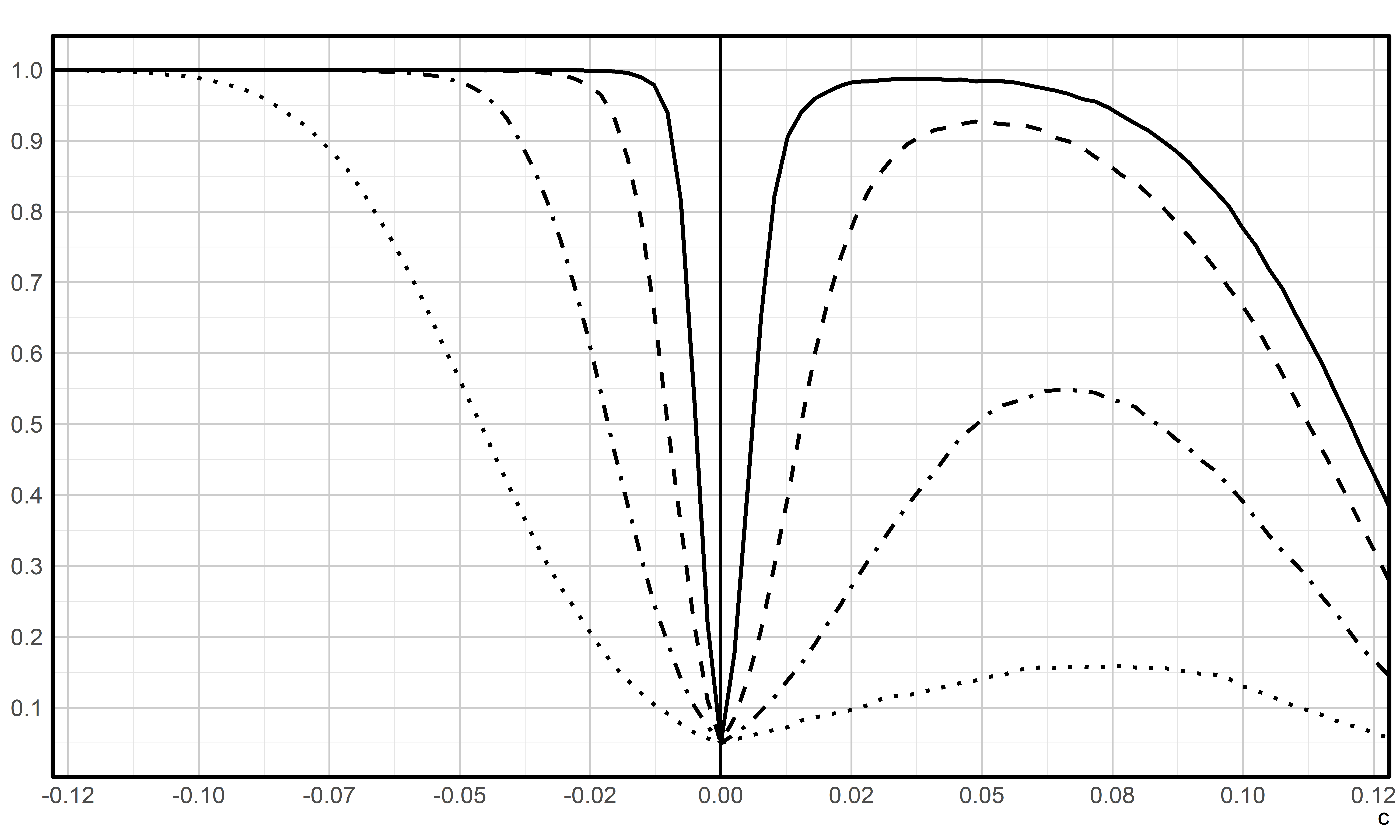}
\end{center}
\caption{\textsc{Rejection frequencies under alternative. }Case of $\alpha
_{0}=0.5$.\ Rejection frequencies for $\tau_{t}>q_{R}$ (right hand side of
$c=0$)\ and $\tau_{t}<q_{L}$(left hand side), with $q_{R}$ [$q_{L}$] the
size-adjusted $0.95$ [$0.05$] quantiles. Solid line: median number of
observations $\operatorname*{med}\{n(t)\}=62500$ for $c=0$; dashed,
dotted-dashed and dotted lines: $\operatorname*{med}\{n(t)\}$ equal to
$12500$, $2500$ and $500$, respectively. Number of Monte Carlo-replications
$M=10000.$}%
\label{RP2}%
\end{figure}

\bigskip

\begin{figure}[ptb]
\begin{center}
\includegraphics[
trim=0.000000in 0.005504in 0.002033in -0.005504in,
height=8.0cm,
width=12.0cm
]{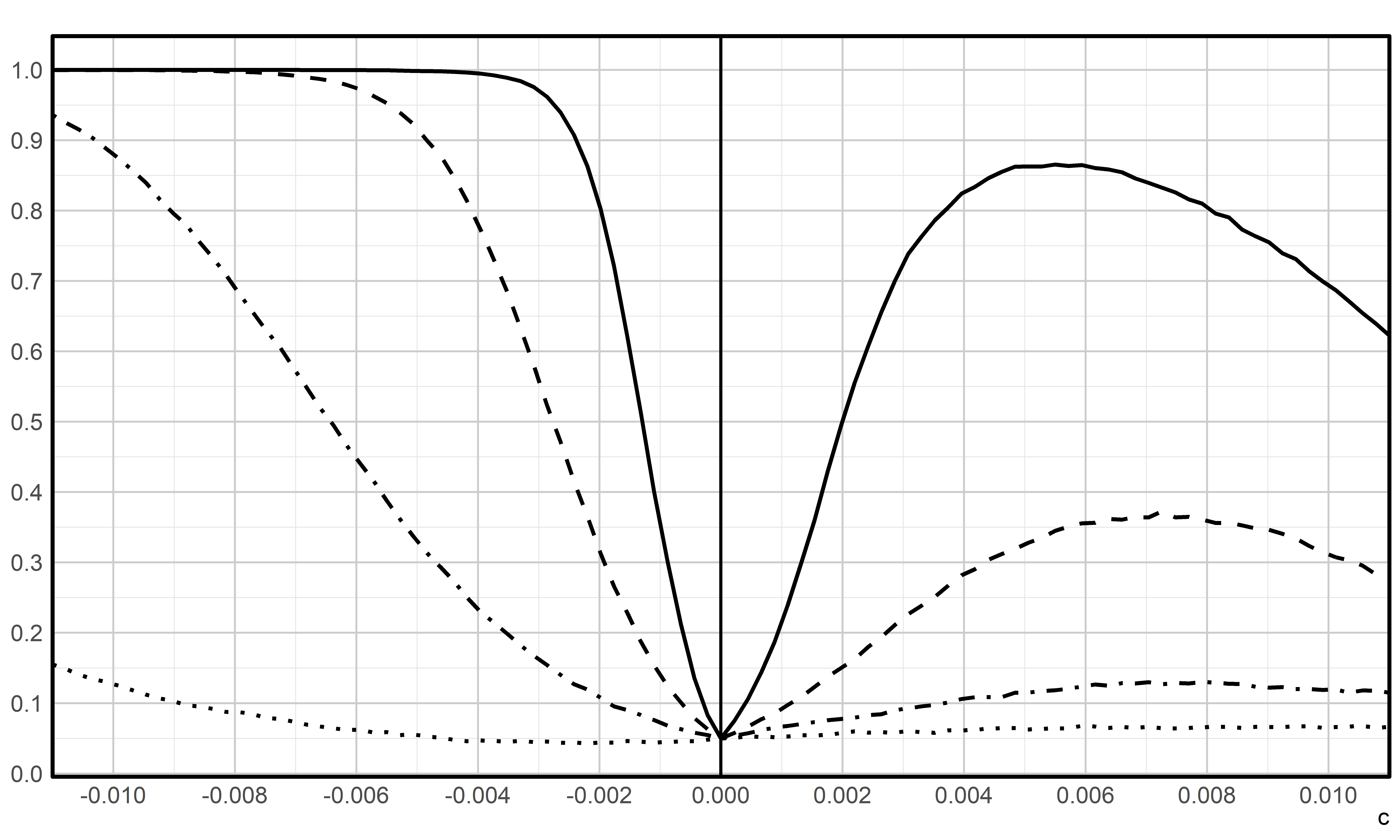}
\end{center}
\caption{\textsc{Rejection frequencies under alternative. }Case of $\alpha
_{0}=0.15$. Rejection frequencies for $\tau_{t}>q_{R}$ (right hand side of
$c=0$)\ and $\tau_{t}<q_{L}$(left hand side), with $q_{R}$ [$q_{L}$] the
size-adjusted $0.95$ [$0.05$] quantiles. Solid line: median number of
observations $\operatorname*{med}\{n(t)\}=62500$ for $c=0$; dashed,
dotted-dashed and dotted lines: $\operatorname*{med}\{n(t)\}$ equal to
$12500$, $2500$ and $500$, respectively. Number of Monte Carlo-replications
$M=10000.$}%
\label{RP3}%
\end{figure}

\subsection{Properties under the alternative}

\label{sec MC power}

To investigate the behavior of the ERPs under the alternative, we focus in
this section on one-sided tests based on $\tau_{t}$; see Section
\ref{ssec testing}.

In terms of parameters $\alpha$ and $\beta$ under the alternatives, we
consider the following design. With $\alpha_{0}$ and $\beta_{0}$ chosen as
under the null, that is, $\alpha_{0}\in\left\{  0.15,0.5,0.85\right\}  $ and
$\beta_{0}=1-\alpha_{0}$, we set
\begin{equation}
\alpha=\alpha_{0}+c\text{, }\beta=1-\alpha_{0}
\label{eq a and b def under alternative}%
\end{equation}
where $c\in I\left(  \alpha_{0}\right)  =[-c\left(  \alpha_{0}\right)
,c\left(  \alpha_{0}\right)  ]$, with $c\left(  \alpha_{0}\right)  >0$ -- and
hence $I\left(  \alpha_{0}\right)  $ -- selected such that the stationarity
condition $\mathbb{E}\left[  \log\left(  \alpha\varepsilon_{i}+\beta\right)
\right]  <0$ holds for all $c\in I\left(  \alpha_{0}\right)  $. Specifically,
with $c\left(  0.15\right)  =0.011$, $c\left(  0.5\right)  =0.128$ and
$c\left(  0.85\right)  =0.149$, we consider an equidistant grid of $p=100$
points in the intervals $I\left(  \alpha_{0}\right)  $.\ In terms of $t$, we
consider different values of $t$ and hence time spans $[0,t]$, such that for
$c=0$ and each $\alpha_{0}$ value, the median number of observations across
Monte Carlo (MC) replications takes values $\operatorname*{med}\left\{
n\left(  t\right)  \right\}  \in\left\{  500,2500,12500,62500\right\}  $.

The corresponding ERPs for $\alpha_{0}=0.85$, $\alpha_{0}=0.50$ and
$\alpha_{0}=0.15$ are reported in Figures \ref{RP1}, \ref{RP2} and \ref{RP3},
respectively. These ERPs are size-adjusted; that is, they are based on
critical values computed using quantiles from the empirical distribution of
the test statistic $\tau_{t}$ under $c=0$.

Consider the (left-sided) test for the null $\alpha+\beta\geq1$, or
$\mathbb{E}[x_{i}]=\infty$ against the finite expectation alternative
$\alpha+\beta<1$; the ERPs for this test correspond to \emph{negative values
}of $c$ in the figures. We note the following facts.

First, as expected, the ERPs increase as $t$ (hence, $\operatorname*{med}%
\left\{  n\left(  t\right)  \right\}  $) gets larger. Second, the ERPs
increase monotonically as $c<0$ moves away from the null of $c=0$. Apart from
the fact that the distance from the null increases in $-c$, this also reflects
that by eq.(8) in CMRV,
\[
n\left(  t\right)  /t\rightarrow_{\text{a.s.}}1/\mathbb{E}[x_{i}]=\left(
1-\left(  \alpha+\beta\right)  \right)  /\omega_{0}=-c\text{,}%
\]
using (\ref{eq a and b def under alternative}). Hence for $c<0$, $n\left(
t\right)  $ is proportional to $\left(  -c\right)  t$, enhancing the observed
increase in ERPs as $-c$ increases. Third, for a given value of $c<0$, the
ERPs increase as $\alpha_{0}$ becomes smaller. For instance, when $c=-0.01$
and $\operatorname*{med}\left\{  n\left(  t\right)  \right\}  =2500$, the ERP
for $\alpha_{0}=0.15$ is close to $90\%$, while for $\alpha_{0}=0.50$
($\alpha_{0}=0.85$) it is approximately $20\%$ ($10\%$). This indicates that
the power of the test is highest in regions of the parameter space associated
with small $\alpha_{0}$. Since small values of $\alpha$ are frequently
encountered in applied work, this suggests that the test performs well where
it is most likely to be used in practice.

Next, consider the right-sided test for the null $\alpha+\beta\leq1$ against
$\alpha+\beta>1$, or $c>0$ in (\ref{eq a and b def under alternative}). As for
the left-sided test, the ERPs increase with $t$. Moreover, for a given value
of $c$, the ERPs are highest when $\alpha_{0}$ attains its smallest value,
i.e. $\alpha_{0}=0.15$. We note that by the implicit definition of $\kappa$
(see (\ref{eq kappa})), it holds that for $\alpha_{1}=1-\beta_{1}<\alpha
_{2}=1-\beta_{2}$, then $\kappa$ as a function of $\alpha_{i}$, $\kappa\left(
\alpha_{i},c\right)  $, $i=1,2$ and $c$ fixed, satisfies $\kappa\left(
\alpha_{1},c\right)  <\kappa\left(  \alpha_{2},c\right)  $. That is, as is
well-known also from the integrated GARCH literature, for smaller values of
$\alpha_{0}$ (and hence larger $\beta_{0}=1-\alpha_{0}$) the tail index varies
more as $c$ is varying, resulting in the larger ERPs. A further notable
difference from the left-sided test is that for a fixed time span $\left[
0,t\right]  $\ the ERPs are not monotone in $c$. To explain this, let
$\theta_{c}=\left(  \omega_{0},\alpha,\beta\right)  $ with $\alpha,\beta$
defined in (\ref{eq a and b def under alternative}) such that $\theta
_{0}=\left(  \omega_{0},\alpha_{0},\beta_{0}\right)  $. It follows that
$\psi_{i}\left(  \theta_{c}\right)  >\psi_{i}\left(  \theta_{0}\right)  $,
with (for the stationary solution)%
\[
\psi_{i}\left(  \theta_{c}\right)  =\omega_{0}\Big[  1+\sum_{j=0}^{\infty
}\prod_{k=0}^{j}\left(  \alpha\varepsilon_{i-1-k}+\beta\right)  \Big]
\text{,}%
\]
implying durations are increasing in $c$ as $x_{i}\left(  c\right)  =\psi
_{i}\left(  \theta_{c}\right)  \varepsilon_{i}>x_{i}=\psi_{i}\left(
\theta_{0}\right)  \varepsilon_{i}.$ That is, the observed number of
observations $n\left(  t\right)  $ is decreasing in $c$, which leads to the
observed loss in rejection probabilities for fixed $\left[  0,t\right]  $. As
noted above, this effect is not present for the left-sided test, where, for
$c<0$, we have the opposite effect: as $c$ decreases, durations decrease and
$n(t)$ increases, leading to an increase in ERPs.

\section{Empirical illustration}

\label{sec empirical}

In the seminal work by Engle and Russell~(1998), the ACD model was applied to
analyze durations between intra-day trades of the IBM stock over a three-month
period. Since then, it has been widely used in applications involving
high-frequency trade-durations for various financial assets; see, e.g.,
Aquilina, Budish and O'Neill~(2022), Hamilton and Jorda~(2002) and Saulo, Pal,
Souza, Vila and Dasilva~(2025).

We illustrate our results by applying ACD models to intra-day,
diurnally-adjusted durations $\left\{  x_{i}\right\}  $ for five different
exchange-traded funds (ETFs) tracking cryptocurrency prices from January 2 to
February 28, 2025 (or, 35 trading days). The ETFs considered are the Grayscale
Bitcoin Mini Trust (ticker: BTC), Grayscale Ethereum Mini Trust (ETH),
Grayscale Bitcoin Trust (GBTC), Grayscale Ethereum Trust (ETHE) and Bitwise
Bitcoin (BITB).

\begin{figure}[t]
\begin{center}
\includegraphics[
trim=0.000000in 0.005504in 0.002033in -0.005504in,
height=8.0cm,
width=12.0cm
]{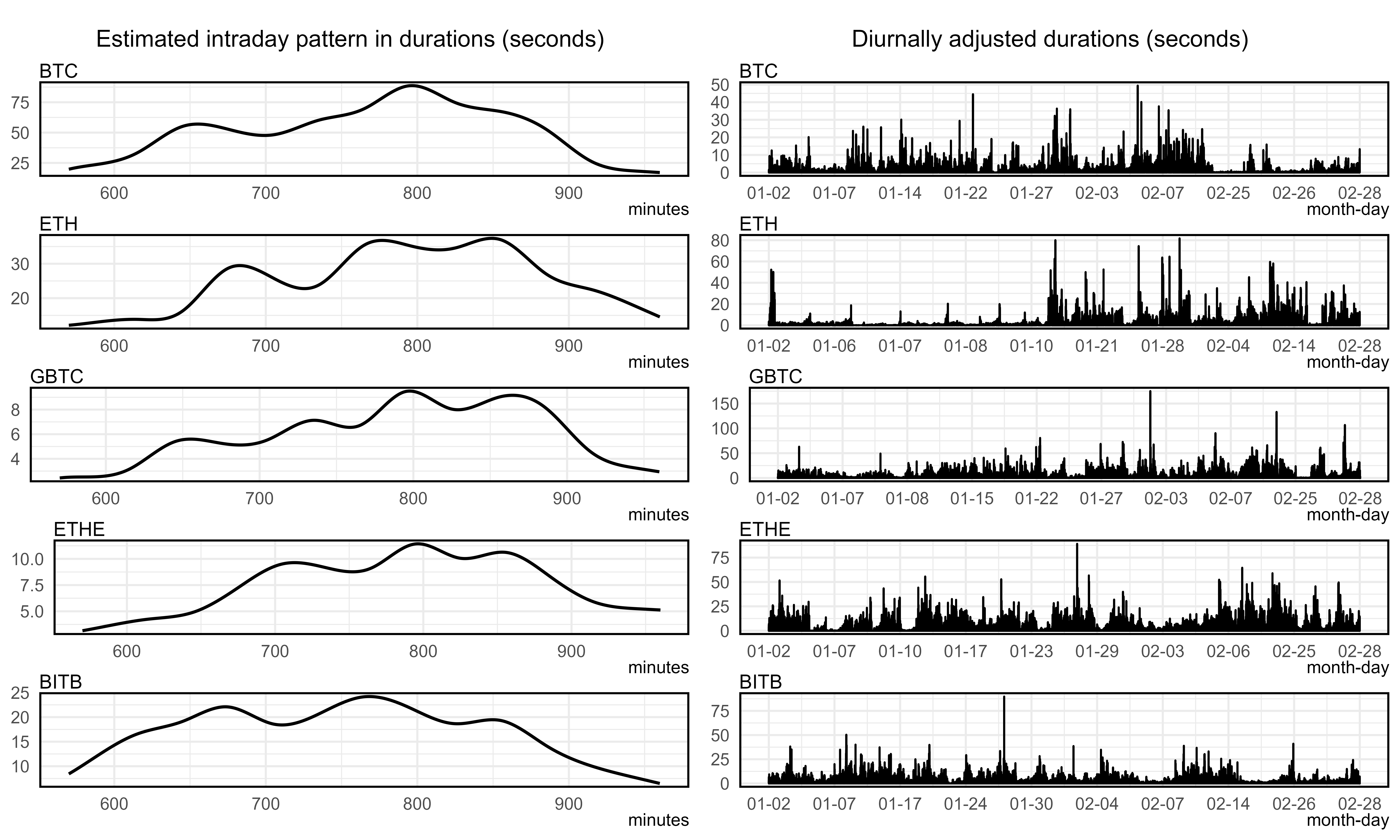}
\end{center}
\caption{\textsc{Durations and Diurnal Pattern.} Right column: Diurnally
adjusted durations $x_{i}$ (in seconds) as a function of calender time. Left
column: Estimated diurnal intra-daily pattern in durations $x_{i}$ as a
function of time (corresponding to 9:30am-4pm). }%
\label{Durations}%
\end{figure}

Intra-day durations for the observed ETFs are measured in seconds (with
decimal precision down to nano-seconds) and are obtained from the limit order
book records on the NASDAQ stock exchange using the LOBSTER database
(\texttt{https://lobsterdata.com/index.php)}. As detailed in Hautsch (2012,
Ch.3), the original, or `raw'\ intra-day durations obtained from the limit
order book are corrected for intradaily patterns using here cubic splines
(with knots placed every 30 minutes). Figure \ref{Durations} shows the
obtained diurnally adjusted durations $\left\{  x_{i}\right\}  $ for each of
the ETFs, together with the estimated intradaily patterns for the different
ETFs;\ as expected, more frequent trading (and hence shorter durations) is
observed at the market open and close, relative to the mid-day period. The
observation period of 35 trading days during regular trading hours (9:30 AM to
4:00 PM EST) corresponds to the time span $[0,t]$, where $t=35\cdot
23400=819,000$ seconds. As to the number of trades $n(t)$ for each of the
ETFs\ these are respectively: $19,366$ for BTC, $35,492$ for ETH, $157,620$
for GBTC, $120,104$ for ETHE, and $n\left(  t\right)  =51,917$ for BITB.
Although the number of trades may appear comparatively low, this reflects
moderate intra-day liquidity exhibited by the ETFs, which is typical of
exchange-traded products and contrasts with the high-frequency trading
activity commonly observed on cryptocurrency exchanges.

For each of the five series, we estimate the ACD model in
(\ref{eq ACD model x}) with QMLEs obtained by maximization of the
log-likelihood function in (\ref{eq def Likelihood}) with initial values
$x_{0}$ and $\psi_{0}\left(  \theta\right)  =x_{0}$. Note that Engle and
Russell (1998) reset the initial value of $\psi_{i}\left(  \theta\right)  $ on
every new trading day; adopting their approach instead of the one used here
yields virtually identical empirical results.

Parameter estimates of $\theta:=(\omega,\alpha,\beta)^{\prime}$ over the
selected time span $[0,t]$, denoted by $\hat{\theta}_{t}:=(\hat{\omega}%
_{t},\hat{\alpha}_{t},\hat{\beta}_{t})^{\prime}$, are reported in Table
\ref{tab param est} along with robust standard errors computed as in
(\ref{eq cons Sigma}). The table also reports the corresponding $t$ statistics
$\tau_{t}$ from (\ref{eq def t stat}) and the quasi-likelihood ratio
statistics $\operatorname{QLR}_{t}$ from (\ref{def QLR stat}) for testing the
null hypothesis $\mathsf{H}_{\text{IACD}}:\alpha+\beta=1$; these statistics
can be used to perform (one-sided or two-sided) tests for the IACD
specification, as well as tests for the null hypothesis of infinite expected
duration, $\mathbb{E}[x_{i}]=\infty$ against the alternative of finite
expected duration, $\mathbb{E}[x_{i}]<\infty$; see the discussion below.

\begin{figure}[h]
\begin{center}
\includegraphics[
trim=0.000000in 0.005504in 0.002033in -0.005504in,
height=8.0cm,
width=12.0cm
]{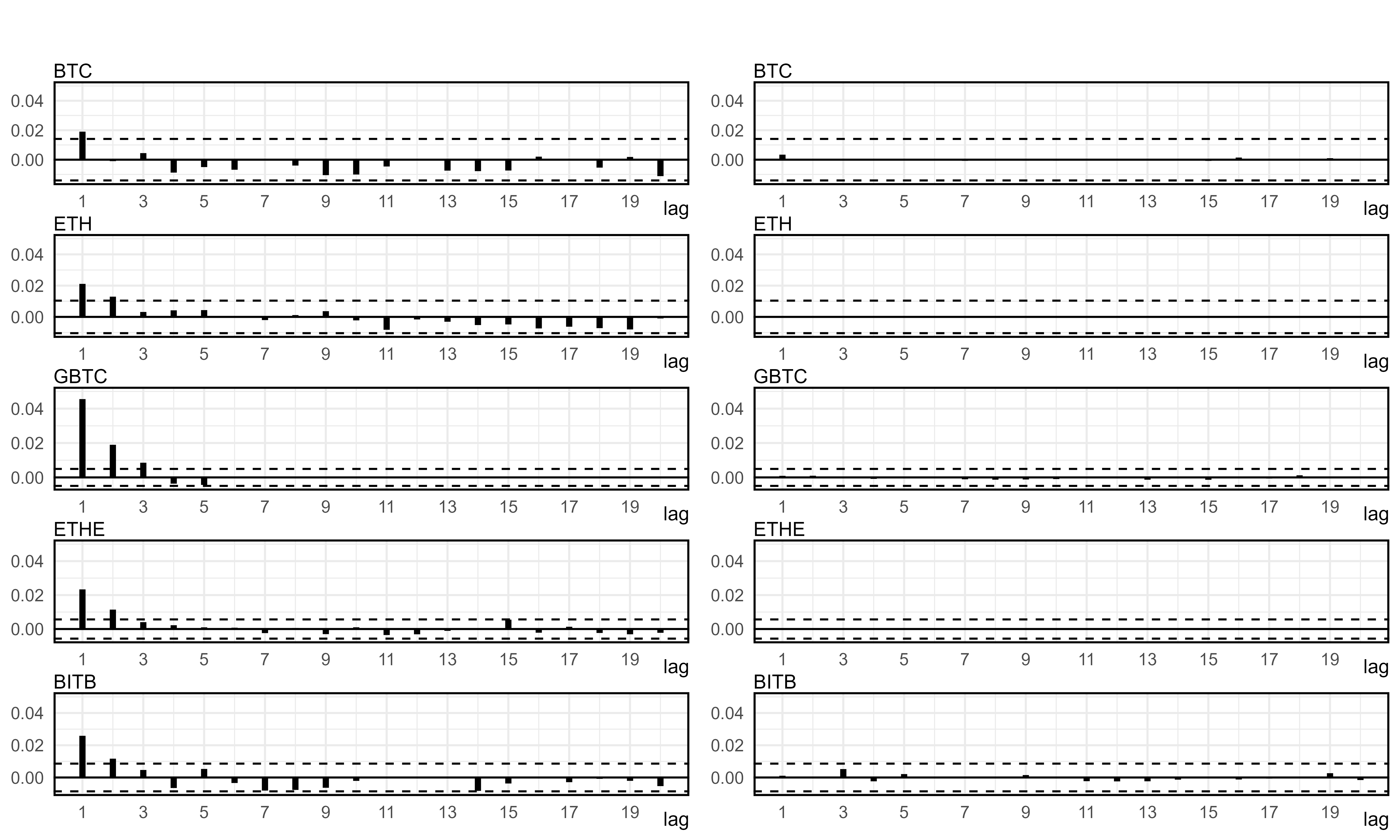}
\end{center}
\caption{\textsc{ACF\ plots.} Sample autocorrelation function (ACF) for the
estimated residuals, $\hat{\varepsilon}_{i}=x_{i}/\psi_{i}(\hat{\theta}_{t})$
(left column) and the squared estimated residuals, $\hat{\varepsilon}_{i}^{2}$
(right column). Plots include (dashes) standard 0.95-confidence intervals.}%
\label{ACF}%
\end{figure}

The model appears to be reasonably well specified for all five series. In
particular, as indicated in Figure \ref{ACF}, some autocorrelation remains in
the standardized residuals, $\hat{\varepsilon}_{i}=x_{i}/\psi_{i}(\hat{\theta
}_{t})$, although their squared values exhibit no significant autocorrelation.
This type of dependence in the residuals $\hat{\varepsilon}_{i}$ is consistent
with previous findings in the ACD literature, where it is well documented that
fully eliminating all serial correlation from the residuals can be
challenging; see e.g. Pacurar (2008). Importantly, the empirical distribution
of the $\hat{\varepsilon}_{i}$'s does not appear to be exponential, again
consistent with findings commonly reported in the financial durations
literature; see, for example, Section 5.3.1 of Hautsch~(2012) and the
references therein.

Returning to the results in Table \ref{tab param est}, we first test the null
hypothesis of infinite expected duration $\mathbb{E}[x_{i}]=\infty$ against
the alternative $\mathbb{E}[x_{i}]<\infty$. This hypothesis can be assessed by
testing the null hypothesis $\alpha+\beta\geq1$ against the one-sided
alternative $\alpha+\beta<1$, using the $t$ statistics $\tau_{t}$ from
(\ref{eq def t stat}). The $\tau_{t}$ statistics are all positive, and thus we
do not reject the null hypothesis of infinite mean ($\mathbb{E}\left[
x_{i}\right]  =\infty$) for any of the five series.

We next consider the null hypothesis of integrated ACD, $\mathsf{H}%
_{\text{IACD}}:\alpha+\beta=1$. This can be tested against the two-sided
alternative, $\alpha+\beta\neq1$, or against one sided alternatives, such as
$\alpha+\beta>1$ or $\alpha+\beta<1$. Using the t-statistic $\tau_{t}$ defined
in (\ref{eq def t stat}) and the $\operatorname{QLR}_{t}$ statistics defined
in (\ref{def QLR stat}), the null hypothesis is rejected at the $5\%$ nominal
level for four out of the five cryptocurrency ETF considered: BTC, ETH, GBTC
and ETHE. The IACD specification is supported for the BITB cryptocurrency ETF only.

\bigskip%

\begin{table}[t] \centering
\caption{\textsc{ACD(1,1) estimates and IACD test statistics}} \vspace{0.5em}%
\begin{tabular}
[c]{lcccccc}\hline
& $\omega$ & $\alpha$ & $\beta$ & $\alpha+\beta$ & $t_{\alpha+\beta=1}$ &
$\operatorname*{QLR}_{t}$\\\hline
BTC & \multicolumn{1}{r}{$\underset{(0.662)}{6.663}$} &
\multicolumn{1}{r}{$\underset{(0.010)}{0.186}$} &
\multicolumn{1}{r}{$\underset{(0.007)}{0.829}$} &
\multicolumn{1}{r}{$\underset{(0.004)}{1.015}$} & \multicolumn{1}{r}{$3.72$} &
\multicolumn{1}{r}{$16.84$}\\
ETH & \multicolumn{1}{r}{$\underset{(0.004)}{0.007}$} &
\multicolumn{1}{r}{$\underset{(0.007)}{0.123}$} &
\multicolumn{1}{r}{$\underset{(0.004)}{0.896}$} &
\multicolumn{1}{r}{$\underset{(0.004)}{1.018}$} & \multicolumn{1}{r}{$4.86$} &
\multicolumn{1}{r}{$76.88$}\\
GBTC & \multicolumn{1}{r}{$\underset{(0.221)}{1.974}$} &
\multicolumn{1}{r}{$\underset{(0.003)}{0.119}$} &
\multicolumn{1}{r}{$\underset{(0.002)}{0.896}$} &
\multicolumn{1}{r}{$\underset{(0.001)}{1.015}$} & \multicolumn{1}{r}{$13.77$}
& \multicolumn{1}{r}{$235.92$}\\
ETHE & \multicolumn{1}{r}{$\underset{(0.041)}{0.394}$} &
\multicolumn{1}{r}{$\underset{(0.003)}{0.083}$} &
\multicolumn{1}{r}{$\underset{(0.002)}{0.927}$} &
\multicolumn{1}{r}{$\underset{(0.001)}{1.010}$} & \multicolumn{1}{r}{$8.27$} &
\multicolumn{1}{r}{$105.90$}\\
BITB & \multicolumn{1}{r}{$\underset{(0.536)}{4.836}$} &
\multicolumn{1}{r}{$\underset{(0.004)}{0.095}$} &
\multicolumn{1}{r}{$\underset{(0.003)}{0.906}$} &
\multicolumn{1}{r}{$\underset{(0.001)}{1.002}$} & \multicolumn{1}{r}{$1.43$} &
\multicolumn{1}{r}{$1.58$}\\\hline
\end{tabular}
$\ $

\begin{flushleft}
{\small \textit{Notes:} Parameter estimates (three decimal points) with
standard errors in parentheses, together with the $\tau_{t}$ and
$\operatorname*{QLR}_{t}$ statistics. Note that ${\hat{\omega}_{t}}$ has been
scaled by ${10^{3}}$. }
\end{flushleft}

\label{tab param est}%
\end{table}%

Taken together, our results show that diurnally adjusted trade durations for
cryptocurrency ETFs are heavy-tailed, with infinite expectation and an implied
tail index $\kappa$ less than (or equal to) one. These findings underscore the
importance of using statistical models that accommodate infinite expected
durations and tail indexes at or below one when analyzing and modeling
high-frequency financial durations in cryptocurrency markets.

\section{Conclusions}

\label{sec conc}

In this paper, we have completed the asymptotic theory for (quasi)
likelihood-based estimation in autoregressive conditional duration (ACD)
models, specifically addressing the previously unresolved `integrated
ACD'\ case where the parameters satisfy the critical condition $\alpha
+\beta=1$. We have established three main results. First, the rate of
convergence of the QML estimators differs from both the $\alpha+\beta>1$ case
and the $\alpha+\beta<1$ the case: interestingly, we find a discontinuity,
with the rate being $\sqrt{t/\log t}$ when $\alpha+\beta=1$, where $t$ denotes
the length of the observation period. Second, despite this nonstandard rate,
the QMLE remains asymptotically Gaussian. Third, standard inference procedures
-- based on $t$-statistics and likelihood ratio tests -- remain valid under
the integrated ACD setting. We have characterized by Monte Carlo simulation
the quality of the asymptotic approximations, finding that empirical rejection
frequencies of tests are close to the selected nominal levels, albeit large
samples are required for some parameter configurations. Finally, we have
applied our results to recent high-frequency trading data on various
cryptocurrency ETFs. The empirical evidence indicates heavy-tailed duration
distributions, and in most cases, the integrated ACD hypothesis is not
rejected in favor of the alternative $\alpha+\beta<1$.

An important extension of our work concerns the development of bootstrap
inference methods within this framework. Bootstrap theory exists for the case
of \emph{deterministic} $n\left(  t\right)  $ as for the class of
multiplicative error (MEM) models (see, e.g., Perera, Hidalgo and Silvapulle,
2016, and Hidalgo and Zaffaroni, 2007), and for point processes, such as ACD
models, with finite expected durations (see, e.g., Cavaliere, Lu, Rahbek and
St\ae rk-\O stergaard, 2023). To the best of our knowledge, no bootstrap
theory currently accommodates cases where $\alpha+\beta\geq1$. This
significant and open research question is currently being investigated by the authors.

\section*{References}

\noindent\textsc{Aquilina, Matteo, Eric Budish, and Peter O'Neill, P.} (2022)
\textquotedblleft Quantifying the High-Frequency Trading Arms
Race,\textquotedblright\ \emph{The Quarterly Journal of Economics}, 137, 493--564.

\medskip 
\noindent\textsc{Berkes, Istvan, Lajos Horvath and Piotr Kokoszka} (2003)
\textquotedblleft GARCH\ Processes: Structure and
Estimation,\textquotedblright\ \emph{Bernoulli }, 9(2), 201--227.

\medskip 
\noindent\textsc{Bhogal, Saranjeet K., and Ramanathan Thekke Variyam }(2019)
\textquotedblleft Conditional Duration Models for Highfrequency Data: A Review
on Recent Developments,\textquotedblright\ \emph{Journal of Economic Surveys,} 33(1),~252--273.

\medskip 
\noindent\textsc{Busch, Thomas} (2005) \textquotedblleft A Robust LR Test for
the GARCH Model,\textquotedblright\ \emph{Economics Letters,} 88, 358--364.

\medskip 
\noindent\textsc{Buraczewski, Dariusz, Ewa Damek, E. and Thomas Mikosch, T.
}(2016)\textsc{\ }\emph{Stochastic Models with Power-Law Tails}. NY: Springer.

\medskip 
\noindent\textsc{Cavaliere, Giuseppe, Ye Lu, Anders Rahbek and Jacob
St\ae rk-\O stergaard }(2023) \textquotedblleft Bootstrap Inference for Hawkes
and General Point Processes,\textquotedblright\ \emph{Journal of
Econometrics}, 235, 133--165.

\medskip 
\noindent\textsc{Cavaliere, Giuseppe, Thomas Mikosch, Anders Rahbek, and
Frederik Vilandt }(2024) \textquotedblleft Tail Behavior of ACD Models and
Consequences for Likelihood-based Estimation,\textquotedblright\ \emph{Journal
of Econometrics}, 238(2), 105613.

\medskip 
\noindent\textsc{Cavaliere, Giuseppe, Thomas Mikosch, Anders Rahbek, and
Frederik Vilandt }(2025) \textquotedblleft A Comment on: \textquotedblleft
Autoregressive Conditional Duration: A New Model for Irregularly Spaced
Transaction Data,\textquotedblright\ \emph{Econometrica}, 93(2), 719-729.

\medskip 
\noindent\textsc{Engle, Robert F. and Jeffry R. Russell }(1998)
\textquotedblleft Autoregressive Conditional Duration: A New Model for
Irregularly Spaced Transaction Data,\textquotedblright\ \emph{Econometrica},
66(5), 1127--1162.

\medskip 
\noindent\textsc{Fernandes, Marcelo, Marcelo C. Medeiros and Alvaro Veiga}
(2016) \textquotedblleft The \mbox{(Semi-)} Parametric Functional Coefficient
Autoregressive Conditional Duration Model,\textquotedblright%
\ \emph{Econometric Reviews}, 35, 1221--1250.

\medskip 
\noindent\textsc{Francq, Christian and Jean-Michel Zakoian }(2019) \emph{GARCH
Models: Structure, Statistical Inference and Financial Applications}.\emph{
}NY: Wiley.

\medskip 
\noindent\textsc{Francq, Christian and Jean-Michel Zakoian }(2022)
\textquotedblleft Testing the existence of moments for GARCH
processes,\textquotedblright\ \emph{Journal of Econometrics, }227, 47--64.

\medskip 
\noindent\textsc{Gourieroux, Christian and Jean-Michel Zako\"{\i}an }(2017)
\textquotedblleft Local Explosion Modelling by Non-Causal
Process,\textquotedblright\ \emph{Journal of the Royal Statistical Society B},
79, 737--756.

\medskip 
\noindent\textsc{Hamilton, James .D., and Oscar Jord\`{a} } (2002)
\textquotedblleft A Model of the Federal Funds Rate Target,\textquotedblright%
\ \emph{Journal of Political Economy,} 110, 1135--1167.

\medskip 
\noindent\textsc{Hautsch, Nikolaus }(2012) \emph{Econometrics of Financial
High-Frequency Data}, Berlin:\ Springer\emph{.}

\medskip 
\noindent\textsc{Hidalgo, Javier and Paolo Zaffaroni} (2007) \textquotedblleft
A goodness-of-fit test for ARCH$\left(  \infty\right)  $
models,\textquotedblright\ \emph{Journal of Econometrics}, 141, 835--875.

\medskip 
\noindent\textsc{Jakubowski, Adam and Zbigniew S. Szewczak }(2021)
\textquotedblleft Truncated Moments of Perpetuities and a New Central Limit
Theorem for GARCH Processes without Kesten's Regularity,\textquotedblright%
\ \emph{Stochastic Processes and their Applications}, 131, 151-171.

\medskip 
\noindent\textsc{Jensen, S\o ren T. and Anders Rahbek }(2004)
\emph{\textquotedblleft}Asymptotic Inference for Nonstationary
GARCH,\emph{\textquotedblright\ Econometric Theory}, 20(6), 1203--1226\emph{.}

\medskip 
\noindent\textsc{Lee, SanG-Won and Bruce E. Hansen} (1994) \textquotedblleft
Asymptotic Theory for the GARCH(1, 1) Quasi-Maximum Likelihood
Estimator\textquotedblright, \emph{Econometric Theory,} 10, 29--52.

\medskip 
\noindent\textsc{Ling, Shiqing} (2004) \textquotedblleft Estimation and
testing stationarity for double-autoregressive models,\textquotedblright%
\ \emph{Journal of the Royal Statistical Society B}, 66, 63--78.

\medskip 
\noindent\textsc{Lumsdaine, Robin L.} (1995) \textquotedblleft Finite-Sample
Properties of the Maximum Likelihood Estimator in GARCH(1,1) and IGARCH(1,1)
Models: A Monte Carlo Investigation,\textquotedblright\ \emph{Journal of
Business \& Economic Statistics} , 13(1), 1-10.

\medskip 
\noindent\textsc{Lumsdaine, Robin L.} (1996) \textquotedblleft Consistency and
Asymptotic Normality of the Quasi-Maximum Likelihood Estimator in IGARCH(1,1)
and Covariance Stationary GARCH(1,1) Models,\textquotedblright%
\ \emph{Econometrica}, 64(3), 575--96.

\medskip 
\noindent\textsc{Pacurar, Maria} (2008) \textquotedblleft Autoregressive
Conditional Duration Models in Finance: a Survey of the Theoretical and
Empirical Literature,\textquotedblright\ \emph{Journal of Economic Surveys},
22, 711--751.

\medskip 
\noindent\textsc{Pedersen, Rasmus S. and Anders Rahbek }(2019)
\textquotedblleft Testing GARCH-X Type Models,\textquotedblright%
\ \emph{Econometric Theory, }35, 1012--1047.

\medskip 
\noindent\textsc{Perera, Indeewara, Javier Hidalgo and Mervyn J. Silvapulle}
(2016) \textquotedblleft A Goodness of-Fit Test for a Class of Autoregressive
Conditional Duration Models,\textquotedblright\ \emph{Econometric Reviews},
35(6), 1111-1141.

\medskip 
\noindent\textsc{Saulo, Helton, Pal Suvra, Souza Rubens, Roberto Vila and Alan
Dasilva} (2025) \textquotedblleft Parametric Quantile Autoregressive
Conditional Duration Models With Application to Intraday Value-at-Risk
Forecasting,\textquotedblright\ \emph{Journal of Forecasting},~44:~589-605.

\newpage

\appendix

\setcounter{section}{1}

\section*{Appendix}

\label{appendix}

\setcounter{equation}{0}

\renewcommand{\theequation}{A.\arabic{equation}}

\subsection{Proof of Lemma \textsc{\ref{lem n div by g}}}

With $s_{n}=\sum_{i=1}^{n}x_{i},$ and $n$ deterministic, we first establish
that
\begin{equation}
\frac{s_{n}}{n\log n}\rightarrow_{p}c_{0}\text{, \ as }n\rightarrow\infty.
\label{eq LLN 2}%
\end{equation}
To see this, note that $s_{n}=s_{1n}+s_{2n},$ with $s_{1n}=\sum_{i=1}^{n}%
\psi_{i}$ and $s_{2n}=\sum_{i=1}^{n}\psi_{i}(\varepsilon_{i}-1)$, and the
result holds by establishing \emph{(i) }$s_{1n}/\left(  n\log n\right)
\rightarrow_{p}c_{0}$ and \emph{(ii)} $s_{2n}/\left(  n\log n\right)
\rightarrow_{p}0$.

It follows by Lemma 4 in CMRV that $\psi_{i}=\psi_{i}(\theta_{0})$ satisfies
the stochastic recurrence equation,%
\begin{equation}
\psi_{i}=\omega_{0}+\left(  \alpha_{0}\varepsilon_{i-1}+1-\alpha_{0}\right)
\psi_{i-1}=A_{i}\psi_{i-1}+B_{i} \label{eq SRE for psi}%
\end{equation}
with $A_{i}=1+\alpha_{0}(\varepsilon_{i-1}-1)$ and $B_{i}=\omega_{0}$.
Moreover, $\psi_{i}$, and hence also $x_{i}=\psi_{i}\varepsilon_{i}$, have
tail index $\kappa_{0}=1>0$, such that in particular,%
\begin{equation}
\mathbb{P(}\psi_{i}>x)\sim c_{0}x^{-1},\text{ \ \ as }x\rightarrow\infty,
\label{tail of psi result}%
\end{equation}
with $c_{0}$ given by (\ref{eq def c0}). Next, by (\ref{tail of psi result})
and L'H\^{o}pital's rule,%
\begin{equation}
\mathbb{E}[\psi_{i}\mathbb{I(}\psi_{i}\leq n)]=\int_{0}^{n}\mathbb{P}\left(
\psi_{i}>x\right)  dx-n\mathbb{P}\left(  \psi_{i}>n\right)  \sim c_{0}%
\log\left(  n\right)  ,\text{ \ as }n\rightarrow\infty.
\label{eq tail of truncated}%
\end{equation}
Using (\ref{eq tail of truncated}) and (12) and (15) in Theorem 1.1 in
Jakubowski and Szewczak (2021) (henceforth JS), it follows by Theorem 2.1 in
JS, that \emph{(i) }holds. For \emph{(ii) }decompose $s_{2n}$ as follows,%
\begin{align*}
s_{2n}  &  =\sum_{i=1}^{n}\psi_{i}\mathbb{I}(\psi_{i}\leq n\log\left(
n\right)  )(\varepsilon_{i}-1)+\sum_{i=1}^{n}\psi_{i}\mathbb{I}(\psi_{i}%
>n\log\left(  n\right)  )(\varepsilon_{i}-1)\\
&  =s_{21n}+s_{22n}.
\end{align*}
{\ For the first term }$s_{21n}$ {we have}%
\[
\mathbb{V}[s_{21n}/(n\log\left(  n\right)  )]=n\,\mathbb{P}(\psi_{i}%
>n\log\left(  n\right)  )\dfrac{\mathbb{V}[\varepsilon_{i}]\mathbb{E}[\psi
_{i}^{2}\mathbb{I}(\psi_{i}\leq n\log\left(  n\right)  )]}{(n\log\left(
n\right)  )^{2}\mathbb{P}(\psi_{i}>n\log\left(  n\right)  )}.
\]
Using (\ref{tail of psi result}), $n\mathbb{P}(\psi_{1}>n\log\left(  n\right)
)\rightarrow0$, while%
\[
\dfrac{\mathbb{E}[\psi_{i}^{2}\mathbb{I}(\psi_{i}\leq n\log\left(  n\right)
)]}{(n\log\left(  n\right)  )^{2}\mathbb{P}(\psi_{i}>n\log\left(  n\right)
)}\sim1,\text{ \ as }n\rightarrow\infty
\]
by Karamata's theorem (see e.g. pages 26-27 in Bingham, Goldie and Teugels,
(1987)). {Hence }$\emph{s}_{21n}\rightarrow0${\ as desired. For the second
term }$s_{22n}${\ we have for any $\delta>0$
\[
\mathbb{P}(|s_{22n}|>\delta n\log\left(  n\right)  )\leq\mathbb{P}\left(
\bigcup_{i=1}^{n}\left\{  \psi_{i}>n\log\left(  n\right)  \right\}  \right)
\leq n\mathbb{P}(\psi_{1}>n\log\left(  n\right)  )\rightarrow0\,,
\]
using (\ref{tail of psi result}) for the convergence. Hence $s_{22n}%
/(n\log\left(  n\right)  )\rightarrow0$ in probability, and }hence \emph{(ii),
}such that (\ref{eq LLN 2}) holds.

Finally, as by definition $n(t)=\max\left\{  k:\sum_{i=1}^{k}x_{i}\leq
t\right\}  $, using (\ref{eq LLN 2}) and $g\left(  t\right)  =t/\log t$,%
\begin{gather*}
\mathbb{P}\left(  n(t)/g\left(  t\right)  \leq z\right)  =\mathbb{P}\left(
n(t)\leq zg\left(  t\right)  \right)  =\mathbb{P}\left(  \sum_{i=1}^{zg\left(
t\right)  }x_{i}\geq t\right) \\
=1-\mathbb{P}\left(  \left(  \frac{\log\left(  zg\left(  t\right)  \right)
}{\log t}\right)  \left(  zg\left(  t\right)  \log\left(  zg\left(  t\right)
\right)  \right)  ^{-1}\sum_{i=1}^{zg\left(  t\right)  }x_{i}<z^{-1}\right) \\
\rightarrow1-\mathbb{I}(c_{0}<z^{-1})=\mathbb{I}(z\geq c_{0}^{-1})\text{,}%
\end{gather*}
which establishes the desired result, $n(t)/g\left(  t\right)  \rightarrow
_{p}c_{0}^{-1}$.\hfill$\square$

\section{Proof of Theorems \ref{thm QMLE U} and \ref{thm QMLE R and QLR}}

\setcounter{equation}{0}

\renewcommand{\theequation}{B.\arabic{equation}}

\subsection{Proof of Theorem \ref{thm QMLE U}}

We apply Lemma 2.1 in CMRV with $T=t$ replaced there by $g\left(  t\right)
:=t/\log t$ and $\mu=c_{0}$. With $\theta=\left(  \theta_{1},\theta_{2}%
,\theta_{3}\right)  ^{\prime}=\left(  \omega,\alpha,\beta\right)  ^{\prime}$,
conditions (C.1)-(C.3) in CMRV hold by the proof of Theorem 2 there\textbf{.}
To see this, for $n$ deterministic, define $\mathcal{L}_{n}(\theta)$ by
replacing $n(t)$ by $n\,$\ in (\ref{eq def Likelihood}), and define
$\mathcal{S}_{n}(\theta)$ and $\mathcal{I}_{n}(\theta)$ similarly. Then, by
CMRV, as $n\rightarrow\infty$,
\[
\frac{1}{\sqrt{n}}\mathcal{S}_{[n\cdot]}(\theta_{0})\rightarrow_{w}%
\Omega_{\mathcal{S}}^{1/2}\mathcal{B}\left(  \cdot\right)  ,\text{ \ }\frac
{1}{n}\mathcal{I}_{n}(\theta_{0})\rightarrow_{\text{a.s}}\Omega_{\mathcal{I}%
},
\]
where $\mathcal{B}\left(  \cdot\right)  $ is a three dimensional Brownian
motion on $[0,\infty)$, $\Omega_{S}=\mathbb{E}\left[  s_{i}(\theta_{0}%
)s_{i}(\theta_{0})^{\prime}\right]  $ and $\Omega_{I}=\mathbb{E}\left[
\iota_{i}(\theta_{0})\right]  $. Moreover, $\sup_{\theta\in\Theta}\left\vert
\frac{1}{n}\partial^{3}\mathcal{L}_{n}(\theta)/\partial\theta_{i}%
\partial\theta_{j}\partial\theta_{k}\right\vert \leq c_{n}\rightarrow
_{\text{a.s.}}c<\infty$, $i,j,k=1,2,3$, and finally, as (C.4) holds by Lemma
\ref{lem n div by g}, we conclude by Lemma 2.1 in CMRV%
\[
\text{\ }\sqrt{g\left(  t\right)  }(\hat{\theta}_{t}-\theta_{0})^{\prime
}\rightarrow_{d}N\left(  0,c_{0}\Omega_{\mathcal{I}}^{-1}\Omega_{\mathcal{S}%
}\Omega_{\mathcal{I}}^{-1}\right)  \text{ as }t\rightarrow\infty\text{,}%
\]
holds. By standard results from GARCH models, see e.g. Jensen and Rahbek
(2004), $\Omega_{\mathcal{S}}=\mathbb{V}\left[  \varepsilon_{i}\right]
\Omega_{\mathcal{I}}$, using that by assumption $\mathbb{E}\left[
\varepsilon_{i}\right]  =1$, and the desired result in Theorem
\ref{thm QMLE U} holds.\hfill$\square$

\subsection{Proof of Theorem \ref{thm QMLE R and QLR}}

The result in (\ref{eq R QMLE}) follows by the proof of Theorem
\ref{thm QMLE U}. Thus with%
\begin{equation}
\gamma=\partial\theta\left(  \phi\right)  /\partial\phi^{\prime}=\left(
\begin{array}
[c]{ccc}%
1 & 0 & 0\\
0 & 1 & -1
\end{array}
\right)  ^{\prime}, \label{def gamma}%
\end{equation}
$\partial\mathcal{L}_{n(t)}\left(  \theta\left(  \phi\right)  \right)
/\partial\phi=\gamma^{\prime}\mathcal{S}_{n(t)}\left(  \theta\left(  \phi
_{0}\right)  \right)  $ and $\partial^{2}\mathcal{L}_{n(t)}\left(
\theta\left(  \phi_{0}\right)  \right)  /\partial\phi\partial\phi^{\prime
}=-\gamma^{\prime}\mathcal{I}_{n(t)}\left(  \theta\left(  \phi_{0}\right)
\right)  \gamma$ by the chain rule. In particular, (C.1) and (C.2) in Lemma
2.1 of CMRV hold as,
\[
\frac{1}{\sqrt{n}}\gamma^{\prime}\mathcal{S}_{[n\cdot]}\left(  \theta\left(
\phi_{0}\right)  \right)  \rightarrow_{w}\gamma^{\prime}\Omega_{\mathcal{S}%
}^{1/2}\mathcal{B}\left(  \cdot\right)  ,\text{ \ }\frac{1}{n}\gamma^{\prime
}\mathcal{I}_{n}(\theta_{0})\gamma\rightarrow_{\text{a.s}}\gamma^{\prime
}\Omega_{\mathcal{I}}\gamma\text{.}%
\]
Similarly for (C.3), while (C.4) as before holds by Lemma \ref{lem n div by g}%
. Hence (\ref{eq R QMLE}) holds. The asymptotic distribution of the
$\operatorname*{QLR}$ statistic follows by arguments as in the proof of Lemma
2.1 in Pedersen and Rahbek, 2019, using the identity $\Omega_{\mathcal{S}%
}=\mathbb{V}\left[  \varepsilon_{i}\right]  \Omega_{\mathcal{I}}$ and
$n\left(  t\right)  \log\left(  t\right)  /t\rightarrow_{p}1/c_{0}$%
.\hfill$\square$

\end{document}